\documentclass[preprint,aps,prd,groupedaddress,showpacs,floatfix,%
nofootinbib]{revtex4}
\usepackage{graphicx}
\usepackage{amsmath}
\usepackage{bm}

\begin{document}

\preprint{ANL-HEP-PR-06-42}
\title{Reconciling the light-cone and nonrelativistic QCD approaches 
to calculating 
\bm{$e^+e^-\to J/\psi + \eta_c$}
}

\author{Geoffrey T.~Bodwin}
\affiliation{
High Energy Physics Division, 
Argonne National Laboratory, 
9700 South Cass Avenue, Argonne, Illinois 60439}

\author{Daekyoung Kang}
\affiliation{
Department of Physics, Korea University, Seoul 136-701, Korea}
\author{Jungil Lee}
\altaffiliation{Visiting faculty, Physics Department, 
Ohio State University, Columbus, Ohio 43210, USA}
\affiliation{
Department of Physics, Korea University, Seoul 136-701, Korea}


\date{\today}
\begin{abstract}
It has been suggested in Ref.~[A.~E.~Bondar and V.~L.~Chernyak, 
Phys.\ Lett.\ B {\bf 612}, 215 (2005)] that the disagreement between 
theoretical calculations and experimental observations for the rate 
for the process $e^+e^-\to J/\psi + \eta_c$ at the $B$ factories 
might be resolved by using the light-cone method to take into account 
the relative momentum of the heavy-quark and antiquark in the 
quarkonia. The light-cone result for the production cross section
in Ref.~[A.~E.~Bondar and V.~L.~Chernyak, 
Phys.\ Lett.\ B {\bf 612}, 215 (2005)] is almost an order of 
magnitude larger than existing NRQCD factorization results. 
We investigate this apparent theoretical discrepancy. 
We compute light-cone distribution functions by making use of 
quarkonium wave functions from the Cornell potential model. 
Our light-cone distribution functions are similar in shape to those of 
Ref.~[A.~E.~Bondar and V.~L.~Chernyak, Phys.\ Lett.\ B {\bf 612}, 
215 (2005)] and yield a similar cross section.
However, when we subtract parts of the light-cone distribution functions
that correspond to corrections of relative-order $\alpha_s$ in the NRQCD
approach, we find that the cross section decreases by about a factor of
three. When we set certain renormalization factors $Z_i$ in the
light-cone calculation equal to unity, we find a further reduction in
the cross section of about a factor of two. The resulting light-cone
cross section is similar in magnitude to the NRQCD factorization cross
sections and shows only a modest enhancement over the light-cone 
cross section in which the relative momentum of the heavy-quark and 
antiquark is neglected.
\end{abstract}
\pacs{12.38.-t, 12.39.Pn, 12.38.Bx, 13.20.Gd, 14.40.Gx}

\maketitle


\section{Introduction
 \label{intro}}

One of the outstanding issues in quantum chromodynamics (QCD) in recent
years is the discrepancy between the theoretical predictions and the
experimental measurements for the exclusive production of $J/\psi +
\eta_c$ in $e^+e^-$ collisions at the $B$ factories. The cross section
times the branching ratio into at least two charged tracks has been
measured by the Belle collaboration to be $25.6\pm 2.8\pm 3.4~\hbox{fb}$
(Ref.~\cite{Abe:2004ww,Pakhlov:2004au}) and by the BaBar collaboration to be
$17.6\pm 2.8 {}^{+1.5}_{-2.1}~\hbox{fb}$ (Ref.~\cite{Aubert:2005tj}). In
contrast, calculations by Braaten and Lee \cite{Braaten:2002fi} and by
Liu, He, and Chao \cite{Liu:2002wq} using the nonrelativistic QCD (NRQCD)
factorization approach \cite{Bodwin:1994jh} at leading order in
$\alpha_s$ yield predictions of $3.78\pm 1.26~\hbox{fb}$ and
$5.5~\hbox{fb}$, respectively.\footnote{The differences in the
predictions arise from different choices for $\alpha_s$ and the
charmonium wave functions at the origin and from the inclusion of QED
effects in the calculation of Ref.~\cite{Braaten:2002fi}.} A similar
disagreement between NRQCD factorization and experiment holds for
production of $\chi_{c0}$ and $\eta_c(2S)$ mesons in conjunction with a
$J/\psi$ meson. A recent calculation of corrections of next-to-leading
order in $\alpha_s$ leads to an enhancement of the theoretical
prediction of about a factor 1.8 (Ref.~\cite{Zhang:2005ch}).
Nevertheless, a substantial discrepancy between theory and experiment
remains.

It has been suggested in several recent papers
\cite{Ma:2004qf,Bondar:2004sv,Braguta:2005kr,Braguta:2006nf} that
predictions for the $J/\psi+\eta_c$ production cross section in the
light-cone formalism might be in better agreement with experiment. In 
the present paper, we focus on the work of Bondar and Chernyak in 
Ref.~\cite{Bondar:2004sv}, henceforth referred to as BC. In BC, it 
is claimed that an enhancement of the cross section occurs in the 
light-cone formalism because the expressions for the cross section 
contain an integration over the light-cone momentum. In contrast, in the 
NRQCD factorization approach, cross sections are computed as an 
expansion in the velocity $v$ of the heavy quark or antiquark in the 
quarkonium rest frame. In the leading order in $v$, the relative 
momentum of the heavy quark and antiquark are neglected, which is
equivalent in the light-cone formalism to approximating the light-cone
distribution amplitude (light-cone wave function) by a delta function at
a light-cone momentum fraction of $1/2$. Effects of the nonzero relative
momentum of the heavy quark and antiquark in NRQCD are taken into
account through terms of higher order in the $v$ expansion.

The NRQCD factorization and light-cone formalisms are both believed to
be valid approximations to QCD for the process $e^+e^-\to J/\psi+\eta_c$
in the limit in which the hard-scattering momentum transfer is much
greater than either the QCD scale $\Lambda_{\textrm{QCD}}$ or the
charm-quark mass $m_c$.\footnote{For a discussion of the current status
of proofs of NRQCD factorization, see
Refs.~\cite{Nayak:2005rw,Nayak:2005rt,Bodwin:2005ec,Nayak:2006fm}.} In
NRQCD factorization, one makes the further assumption that the cross
section can be written as an expansion in powers of $v$. Since the
light-cone and NRQCD factorization approaches are both first-principles
methods, their predictions should be in reasonable agreement unless, as
is asserted in BC, the $v$ expansion of NRQCD breaks down for the process
$e^+e^-\to J/\psi+\eta_c$.

In this paper we investigate several issues with regard to the
calculation in BC by making use of light-cone distribution amplitudes
that are derived from potential-model quarkonium wave functions. We
first compare the light-cone distribution amplitude that we derive with
the model light-cone distribution amplitudes that are used in BC. We
find that they are quite similar in shape, but with significantly
different asymptotic behavior at large light-cone momentum fraction, and
that they lead to similar values for the cross section for $e^+e^-\to
J/\psi+\eta_c$. We then argue that a part of the contribution to the
cross section that is computed in BC comes from the high-momentum tail
of the light-cone distribution amplitude. Such high-momentum
contributions are taken into account in NRQCD through contributions to
the cross section of relative order $\alpha_s$ and higher. Furthermore,
since our potential-model wave function is accurate only for momenta
that are much less than $m_c$, it is not legitimate for us to include
the high-momentum contribution in our light-cone calculation. When we
subtract this contribution, we find that the cross section is reduced by
about a factor of three. A further reduction of about a factor of two
occurs when we replace with unity certain renormalization factors
$Z_i$ that appear in the calculation in BC, but which have no
counterpart in a conventional NRQCD factorization calculation. With
these modifications, the light-cone calculation gives a cross section
that is comparable to that of the NRQCD factorization calculations.

The remainder of this paper is organized as follows. In
Sec.~\ref{sec:potential-model}, we describe the Cornell potential model
that we use. In Sec.~\ref{sec:LC-BS-wf}, we present the relationships of
the required light-cone distribution amplitudes to the Bethe-Salpeter
(BS) wave function, use the BS equation to compute the BS wave function
in terms of a potential-model wave function, and give the
correspondences between the terms in the resulting BS wave function and
the required light-cone distribution amplitudes. Sec.~\ref{sec:light-cone}
contains the detailed expressions for the light-cone distribution amplitudes
in terms of the potential-model wave function and a
re-arrangement of these expressions into forms that are amenable to
numerical calculation. In Sec.~\ref{sec:pert-corr}, we identify the
high-momentum part of the light-cone distribution amplitude that is
properly treated as an order-$\alpha_s$ correction to the cross section.
We also compute and compare the asymptotic forms of the light-cone
distribution amplitude and the high-momentum part in the
high-momentum limit. In Sec.~\ref{sec:cross-sections}, we present the
expression for the cross section in the light-cone formalism, give our
numerical results for the cross section at $B$-factory energies, and
discuss the implications of these results. Finally, in
Sec.~\ref{sec:summary}, we summarize our findings. 

\section{Potential model}\label{sec:potential-model}
In this section, we describe the calculation of the potential-model
wave function. We note that, if we knew the heavy-quark potential 
exactly, then we could calculate the heavy-quarkonium wave function in a 
potential model, up to corrections of relative order $v^2$ 
(Ref.~\cite{Brambilla:1999xf}). 
We make use of the Cornell potential model of
Ref.~\cite{Eichten:1978tg}. The Cornell potential provides a reasonably
good fit to heavy-quark potentials that are measured in lattice
calculations.\footnote{For a recent review that discusses heavy-quark
potentials from lattice measurements, see Ref.~\cite{Bali:2000gf}.}

The Schr\"odinger equation for the radial wave function for a
quark-antiquark ($Q\bar{Q}$) pair with orbital-angular-momentum
quantum number $\ell$ in a central potential $V(r)$ is
\begin{equation}
\left[
-\frac{1}{mr^2} \frac{d}{dr}\left( r^2 \frac{d}{dr}\right)
+\frac{\ell(\ell+1)}{m r^2}
+V(r)
\right]R(r)=\epsilon_{\textrm{B}} R(r),
\label{radial}
\end{equation}
where $\epsilon_{\textrm{B}}$ is the binding energy, $r$ is the distance 
between the heavy quark and antiquark, and  $m$ is mass of the heavy quark.
Note that the reduced mass of the pair is $\mu=m/2$. 

The Cornell potential \cite{Eichten:1978tg} is given by
\begin{equation}
V(r)=-\frac{\kappa}{r}+\frac{r}{a^2},
\label{model-V}
\end{equation}
where the parameters $\kappa$ and $a$ determine the strength of 
Coulomb and linear potentials, respectively.
For a color-singlet $Q\bar{Q}$ pair, the Coulomb-strength
parameter $\kappa$ can be expressed in terms of an effective
strong coupling $\alpha_s$ as
\begin{equation}
\kappa=\alpha_s C_F,
\label{kappa}
\end{equation}
where $C_F=4/3$.
The parameter $a$ is related to the string tension $\sigma$ as
\begin{equation}
\sigma=\frac{1}{a^2}.
\label{sigma-par}
\end{equation}

Following Ref.~\cite{Eichten:1978tg}, we replace $\kappa$ and $r$ by
dimensionless variables $\lambda$ and $\rho$:
\begin{subequations}
\begin{eqnarray}
\kappa&=&(ma)^{-2/3}\,\lambda,
\label{kappa-lambda}
\\
r&=&\frac{\lambda}{m\kappa}\,\rho= a(ma)^{-1/3}\,\rho.
\label{r-rho}
\end{eqnarray}
\end{subequations}
Substituting the Cornell potential into Eq.~(\ref{radial}), we
can rewrite the radial equation:
\begin{equation}
\left[
\frac{d^2}{d\rho^2}-\frac{\ell(\ell+1)}{\rho^2} 
+\frac{\lambda}{\rho}-\rho+\zeta
\right]u(\rho)=0,
\label{u-eq}
\end{equation}
where the dimensionless function $u(\rho)$ and dimensionless energy
eigenvalue $\zeta$ in Eq.~(\ref{u-eq}) are related to their
physical counterparts as
\begin{subequations}
\label{eq:psir}
\begin{eqnarray}
R(r)&=&\sqrt{\frac{m}{a^2}}\frac{u(\rho)}{\rho},
\\
\epsilon_{\textrm{B}}&=& m (ma)^{-4/3}\zeta.
\end{eqnarray}
\end{subequations}
The wave functions are normalized according to
\begin{eqnarray}
\int_0^\infty |u(\rho)|^2 d\rho=
\int_0^\infty |R(r)|^2 r^2 dr=1.
\end{eqnarray}

In Eq.~(2.19) of Ref.~\cite{Eichten:1978tg}, 
the wave function at the origin is expressed in terms of the model 
parameters $m$ and $a$ as
\begin{equation}
|\psi(0)|^2=\frac{m}{4\pi a^2}
\left(1+\lambda \langle \rho^{-2}\rangle\right),
\label{psi0-ma}
\end{equation}
where the expectation value $\langle \rho^{-2}\rangle$ is defined by
\begin{equation}
 \langle \rho^{-2}\rangle=
\int_0^\infty \frac{\left[u(\rho)\right]^2}{\rho^2}d\rho.
\end{equation}
Numerical values of $\langle \rho^{-2}\rangle$ for various values 
of $\lambda$ are given in Table~I of  Ref.~\cite{Eichten:1978tg}.
As we explain below, we fix the value of $|\psi(0)|$ by using 
the leptonic decay width of the $J/\psi$. We need an additional 
constraint to fix both of the parameters $m$ and $a$ for a given $\lambda$.
Following Ref.~\cite{Eichten:1978tg}, we use the mass splitting 
between the $J/\psi$ and the $\psi(2S)$. Eq.~(2.12) of 
Ref.~\cite{Eichten:1978tg} expresses the mass splitting as

\begin{equation}
M_{\psi(2S)}-M_{J/\psi}=m(ma)^{-4/3}(\zeta_{20}-\zeta_{10}),
\label{M-ma}
\end{equation}
where $\zeta_{n0}$ is the eigenvalue for the radial equation
(\ref{u-eq}) for the principal quantum number $n$ and the
orbital-angular-momentum quantum number $\ell=0$. The values for
$\zeta_{10}$ as a function of $\lambda$ are given in Table~I of
Ref.~\cite{Eichten:1978tg}. We use Eq.~(A4) and Table~II of
Ref.~\cite{Eichten:1978tg} to determine $\zeta_{20}$ for various
values of $\lambda$:
\begin{equation}
\zeta_{20}=4.0879-0.5826\lambda-0.0302\lambda^2.
\label{zeta20}
\end{equation}
Solving Eqs.~(\ref{psi0-ma}) 
and (\ref{M-ma}), we can determine 
the model parameters $m$ and $a$:
\begin{subequations}
\label{m-a}
\begin{eqnarray}
m&=&
\frac{\zeta_{20}-\zeta_{10}}{M_{\psi(2S)}-M_{J/\psi}}
\left( \frac{4\pi|\psi(0)|^2}{1+\lambda \langle \rho^{-2}\rangle} 
\right)^{2/3},
\\
a&=&
\left(\frac{\zeta_{20}-\zeta_{10}}{M_{\psi(2S)}-M_{J/\psi}}\right)^{1/2}
\left( \frac{4\pi|\psi(0)|^2}{1+\lambda \langle \rho^{-2}\rangle} 
\right)^{-1/6}.
\end{eqnarray}
\end{subequations}
From Eqs.~(\ref{kappa}) and (\ref{kappa-lambda}), it can be
seen that the effective strong coupling $\alpha_s$ for the bound-state
potential as a function of $\lambda$ is given by
\begin{equation}
\alpha_s=\frac{\lambda}{C_F}(ma)^{-2/3}.
\label{as}
\end{equation}

As we have mentioned, we determine $|\psi(0)|$ from the leptonic decay width
of the $J/\psi$. NRQCD matrix elements for $J/\psi$ that correspond to
$|\psi(0)|^2$ are given in Table~I of Ref.~\cite{Braaten:2002fi}. These
values are $\langle O\rangle_S=0.208$~GeV$^{3}$ for LO and
$0.335$~GeV$^{3}$ for NLO, where the identifiers LO and NLO indicate
that the short-distance coefficient for the leptonic decay rate of 
the $J/\psi$ has been computed to leading order or next-to-leading order 
in the strong-coupling constant. Expressing $|\psi(0)|$ in terms of these 
matrix elements, we obtain
\begin{equation}
|\psi(0)|=\sqrt{\frac{\langle O\rangle_{J/\psi}}{2N_c}}=
\left\{
\begin{array}{lcl}
0.18619~\textrm{GeV}^{3/2}&& \textrm{(LO)},
\\
0.23629~\textrm{GeV}^{3/2}&& \textrm{(NLO)}.
\end{array}
\right.
\label{psi0-value}
\end{equation}
In this paper, we use the LO value of $|\psi(0)|$, which corresponds
to the formula
\begin{equation}
\Gamma[J/\psi\to \ell^+\ell^-]
=\frac{4\pi e_c^2 \alpha^2}{m_c^2}|\psi(0)|^2,
\end{equation}
where $e_c=2/3$ is the fractional electric charge of the charm quark.
If one uses the NLO value of $|\psi(0)|$, then the average 
momentum-squared of the wave function is reduced by about 10\% 
(Ref.~\cite{BKL}). Therefore, we do not expect that the use of the 
NLO value would change our results qualitatively.

Using $|\psi(0)|=0.18619~$GeV$^{3/2}$, $M_{J/\psi}=3.096916$~GeV,
$M_{\psi(2S)}=3.686093$~GeV, Table~I of Ref.~\cite{Eichten:1978tg}, and
Eq.~(\ref{zeta20}), we can compute $m$ and $a$ from Eq.~(\ref{m-a}). The
results for various values of $\lambda$ are shown in
Table~\ref{table:ma}, along with values for $\alpha_s$ from
Eq.~(\ref{as}), $\sigma$ from Eq.~(\ref{sigma-par}), and
$\gamma_{\textrm{C}}$, which is related to the binding energy for a pure
Coulomb potential and is defined by
\begin{equation}
\gamma_{\textrm{C}}=\frac{1}{2}m \alpha_s C_F.
\label{gamma-C}
\end{equation}
 \begin{table}
 \caption{\label{table:ma}
Potential-model parameters and derived quantities as a function of the
strength $\lambda$ of the Coulomb potential. The definitions of the
parameters and derived quantities are given in the text. The parameters
are computed using the inputs $|\psi(0)|=0.18619$~GeV$^{3/2}$,
$M_{J/\psi}=3.096916$~GeV, and $M_{\psi(2S)}=3.686093$~GeV, as is 
described in the text.
}
 \begin{ruledtabular}
 \begin{tabular}{c|lllllllll}
$\lambda$& 
0& 0.2     & 0.4     & 0.6     & 0.7     & 0.8    & 1.0   & 1.2     & 1.4\\
\hline
$1+\lambda  \langle \rho^{-2}\rangle$&
1.     & 1.24988& 1.55768& 1.93498& 2.15369& 2.39480& 2.95190& 3.62272& 4.42524
\\
$\zeta_{10}$&
2.33811& 2.16732& 1.98850& 1.80107& 1.70394& 1.60441& 1.39788& 1.18084& 0.95264
\\
$\zeta_{20}$&
4.08790& 3.97017& 3.85003& 3.72747& 3.66528& 3.60249& 3.47510& 3.34529& 3.21307
\\
$m$~(GeV)&
1.70670& 1.51548& 1.35120& 1.21003& 1.14710& 1.08877& 0.98458& 0.89501& 0.81796
\\
$a$~(GeV$^{-1}$)&
1.97932& 2.08520& 2.19805& 2.31833& 2.38139& 2.44648& 2.58295& 2.72816& 2.88253
\\
$\sqrt{\sigma}$~(GeV)&
0.50522& 0.47957& 0.45495& 0.43134& 0.41992& 0.40875& 0.38715& 0.36655& 0.34692
\\
$\sigma$~(GeV$^2$)&
0.25525& 0.22999& 0.20698& 0.18606& 0.17633& 0.16708& 0.14989& 0.13436& 0.12035
\\
$\alpha_s$&
0.     & 0.06966& 0.14519& 0.22624& 0.26866& 0.31225& 0.40255& 0.49634& 0.59272
\\
$\gamma_{\textrm{C}}$~(GeV)&
0.     & 0.07037& 0.13079& 0.18250& 0.20545& 0.22664& 0.26423& 0.29615& 0.32322
 \end{tabular}
 \end{ruledtabular}
 \end{table}

Lattice measurements of the heavy-quark potential yield values for
effective coupling $\alpha_s$ of 0.22 in the quenched case and
approximately 0.26 in the unquenched case \cite{Bali:2000gf}. A lattice
measurement of the string tension $K=\sigma$ (Ref.~\cite{Booth:1992bm})
gives $Ka_{\rm L}^2=0.0114(2)$ at a lattice coupling $\beta=6.5$, where
$a_{\rm L}$ is the lattice spacing. Lattice calculations of the hadron
spectrum at $\beta=6.5$ yield values for $1/a_{\rm L}$ of
$3.962(127)$~GeV (Refs.~\cite{Gupta:1996sa,Kim:1993gc}) and
$3.811(59)$~GeV (Refs.~\cite{Gupta:1996sa,Kim:1996cz}). These yield
values of the string tension of $K=0.1790\pm 0.0119$~GeV$^2$ and 
$K=0.1656\pm 0.0059$~GeV$^2$, respectively. 
Comparing the results of these lattice
measurements with Table~\ref{table:ma}, we conclude that $\lambda=0.7$
is a reasonable choice for the value of the potential-model parameter,
and we use that value in our analysis.

\begin{figure}[t]
\includegraphics[width=12cm]{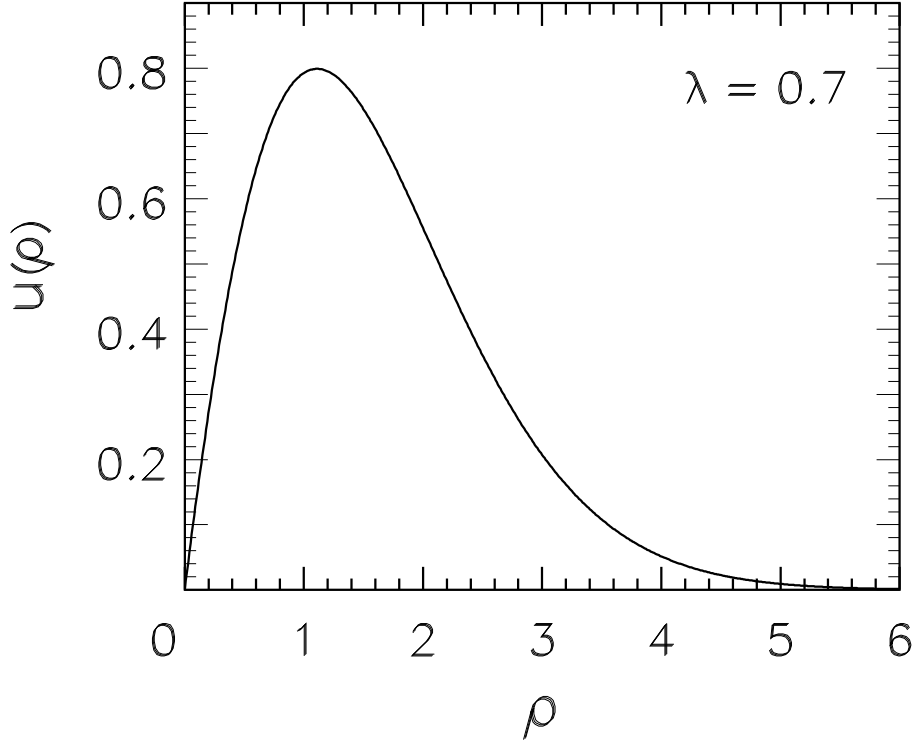}
\caption{
The dimensionless wave function $u(\rho)$ as a function of the
dimensionless variable $\rho$. Input parameters are  $m=1.14710$~GeV,
$a=2.38139$~GeV$^{-1}$, and $\lambda=0.7$.  The wave function at the
origin is taken to be $|\psi(0)|=0.18619~\textrm{GeV}^{3/2}$,
which is the value for the $J/\psi$ wave function at the origin that is 
designated as LO in Ref.~\cite{Braaten:2002fi}.
\label{fig:u}}
\end{figure}

We express the radial equation (\ref{u-eq}) as a difference equation, 
and integrate it numerically. The result, for the choice of parameters 
in Table~\ref{table:ma} that corresponds to $\lambda=0.7$ is shown in 
Fig.~\ref{fig:u}.

\section{Relationships of the light-cone distribution amplitudes to the 
potential-model wave function}\label{sec:LC-BS-wf}
\subsection{Light-cone distribution amplitudes}
In this subsection, we re-produce the expressions that are given in BC
for the light-cone distribution amplitudes in terms of the BS wave functions.

For the $J/\psi$ with polarization $\epsilon$, the light-cone 
distribution amplitudes are defined by
\begin{eqnarray}
\langle J/\psi (P,\epsilon)|
\bar{Q}_{\beta}(x)\,Q_{\alpha}(-x)|0\rangle_{\mu^*}
&=&
\frac{f_{V}M_{J/\psi}}{4}\int^1_0 dz_1\,e^{iP\cdot x(z_1-z_2)}
\biggl \{ /\!\!\!\epsilon\,V_{\perp}(z)
+{/\!\!\!\! P}\,\frac{\epsilon\cdot x}{P\cdot x}\,{\widetilde V}(z)
\nonumber\\
&&\hbox{}\hskip -3cm +f^{t}_{v}(\mu^*) 
    \frac{\sigma_{\mu\nu} \epsilon^{\mu}\, P^{\nu}}{M_{J/\psi}}\,V_{T}(z)
     + f^{a}_{v}(\mu^*)\epsilon_{\mu\nu\rho\sigma}\gamma^{\mu}\gamma_5\,
                        \epsilon^\nu\,P^{\rho}x^{\sigma}\, V_{A}(z) 
\biggl \}_{\alpha\beta}.
\label{bs-lc}
\end{eqnarray}
Here 
$Q$ and $\bar Q$ are the heavy quark and antiquark fields, respectively, 
$\mu^*$ is the renormalization scale, 
$\sigma_{\mu\nu}=\frac{1}{2}[\gamma_{\mu},\gamma_{\nu}]$, and
$z_1=z$ and $z_2=1-z$ are the fractions of the meson momentum
$P^+=P^0+P^3\equiv q_0$ carried by the quark and antiquark,
respectively, at large $P^3$. In deriving these expressions, the
authors of BC neglect the quark transverse momentum inside the heavy
quarkonium in comparison with the quark mass. This is a further
approximation that goes beyond the standard light-cone formalism.
As is pointed out in BC, the light-cone distribution amplitude 
${\widetilde V}(z)$ 
may be eliminated in favor of the standard leading-twist wave
function $V_L(z)$ of a longitudinally polarized $J/\psi$
(Ref.~\cite{Chernyak:1983ej}):
\begin{equation}
\langle J/\psi_{\lambda=0}(P)|\bar{Q}(x)\gamma_{\mu}Q(-x)|0\rangle 
=f_{V}P_\mu
\int^1_0 dz_1\,e^{i P\cdot x(z_1-z_2)}V_{L}(z). 
\label{bs-lc2}
\end{equation}
Taking $\epsilon$ to correspond to helicity $\lambda=0$  in
Eq.~(\ref{bs-lc}) and using
$M_{J/\psi}\epsilon^{\mu}_{\lambda=0}\rightarrow P^{\mu}$ for large $P$,
one obtains
\begin{equation}
V_{L}(z)=V_{\perp}(z)+\widetilde{V}(z).
\label{V-tilde}
\end{equation}

For the $\eta_c$-meson, one has 
\begin{eqnarray}
\langle \eta_c(P)|\bar{Q}_{\beta}(x)\,Q_{\alpha}(-x)|0\rangle_{\mu^*}
&=&
i\frac{f_{P}M_{\eta_c}}{4}
\int^1_0 dz_1 e^{i P\cdot x(z_1-z_2)}\biggl \{ 
\frac{/\!\!\!\! P\,\gamma_5}{M_{\eta_c}}\,P_{A}(z)
-f_{p}^{p}(\mu^*)\,\gamma_5\,P_{P}(z)
\nonumber\\
&&\quad+f_{p}^t(\mu^*)\, 
\sigma_{\mu\nu}\,P^{\mu}\, x^{\nu}\,P_{T}(z)\biggr \}_{\alpha\beta} \,.
\label{bs-lc3}
\end{eqnarray}

\subsection{Relationships to potential-model wave functions}

In order to express the light-cone distribution amplitude in terms of a
potential-model wave function, we now derive an approximate form for the
BS wave function in terms of a potential-model wave function. Here we 
make explicit use of an expansion in the velocity $v$ of the $Q$ or 
$\bar Q$ in the quarkonium rest frame. Of course, such an expansion 
is not valid at large $Q\bar{Q}$ relative momenta. However, as we will 
discuss in detail in Sec.~\ref{sec:pert-corr}, we ultimately subtract such 
large-momentum contributions in the cross-section calculation.

Consider a heavy $Q\bar{Q}$ pair bound in a potential
$V$. The four-momenta for the pair can be written in terms 
of quarkonium momentum $P$ and their relative momentum $p$ as
\begin{subequations}
\label{pQ}
\begin{eqnarray}
p_Q&=&\frac{1}{2}P+p,
\\
p_{\bar{Q}}&=&\frac{1}{2}P-p.
\end{eqnarray}
\end{subequations}
In the rest frame of the heavy quarkonium, $P$ and $p$ are given by
\begin{subequations}
\label{Pp}
\begin{eqnarray}
P&=&(M_{\rm meson},\bm{0})=(2m+\epsilon_\textrm{B},\bm{0}),
\label{P}
\\
p&=&(p^0,\bm{p}),
\label{p}
\end{eqnarray}
\end{subequations}
where $\epsilon_{\textrm{B}}$ is the binding energy of the quarkonium.
Let $\psi(\bm{x})$ be the potential-model wave function satisfying
the time-independent Schr\"odinger equation:
\begin{equation}
\left(\epsilon_{\textrm{B}}-
\frac{\hat{\bm{p}}^2}{2\mu}
\right)\psi(\bm{x})
=
V(\bm{x})
\psi(\bm{x}),
\end{equation}
where $\mu=m/2$ is the reduced mass and $\hat{\bm{p}}$ is the momentum
operator. We specialize to the $S$-wave case. Then
$\psi(\bm{x})=\psi(r)$, where $r=|\bm{x}|$. 

We can form a trial BS wave function, accurate to leading order in $v$, 
by appending spin wave functions onto $\psi$. In the rest
frame of the quarkonium, the spin-singlet and spin-triplet
wave functions are
\begin{subequations}
\label{spin-projectors}
\begin{eqnarray}
\label{proj-1}
(\Pi_0)_{ab}= \sum_{i,j} c_{ij}({}^1S_0) \xi^{i}_{a}\bar{\eta}^{j}_{b}
&=&
\frac{1}{2\sqrt{2}}
\left[
   \left(1+\gamma^0\right)
\gamma^5
\right]_{ab}
=\frac{1}{2\sqrt{2}}
\left[
   \left(1+\frac{/\!\!\!\! P}{2m}\right)
\gamma^5
\right]_{ab},
\\
\label{proj-3}
(\Pi_1)_{ab}= \sum_{i,j} c_{ij}({}^3S_1) \xi^{i}_{a}\bar{\eta}^{j}_{b}
&=&
-\frac{1}{2\sqrt{2}}
\left[
   \left(1+\gamma^0\right)
/\!\!\! \epsilon
\right]_{ab}
=-\frac{1}{2\sqrt{2}}
\left[
   \left(1+\frac{/\!\!\!\! P}{2m}\right)
/\!\!\! \epsilon
\right]_{ab},
\end{eqnarray}
\end{subequations}
where $\xi$ and $\bar{\eta}$ are Dirac spinors for a static free quark
and antiquark, respectively. $c_{ij}({}^1S_0)$ and $c_{ij}({}^3S_1)$
are Clebsch-Gordan coefficients of the spin-singlet and spin-triplet
states for the spin states $i$ and $j$ of $Q$ and $\bar{Q}$,
respectively, and $a$ and $b$ are the Dirac indices. SU(3) color indices
are suppressed. In the last equalities of
Eqs.~(\ref{spin-projectors}), we have written the expressions in a
Lorentz-invariant form.\footnote{These Lorentz-invariant expressions
were first given in Refs.~\cite{Kuhn:1979bb,Guberina:1980dc}. Exact
expressions, correct to all orders in $v$ were given in
Ref.~\cite{Bodwin:2002hg}.} Our trial wave function is then $\psi
(r)(\Pi_\lambda)_{ab}$, where $\lambda$ takes on the values $0$ or $1$. 
In momentum space, the trial
BS wave function is
\begin{equation}
\widetilde{\psi}(\bm{p})2\pi\delta(p^0)(\Pi_\lambda)_{ab}=
\int d^4x\,e^{-ip\cdot x}\psi(r)(\Pi_\lambda)_{ab}.
\label{trial-wf}
\end{equation}

The momentum-space form of the BS equation 
\cite{Salpeter:1951sz} is
\begin{equation}
\label{BS1}
\widetilde{\Psi}^{\textrm{BS}}_{ab}(p)=
\int \frac{d^4q}{(2\pi)^4}\, \left[S^{Q}_{F}(p)\right]_{aa'}
\left[K(p-q)\right]_{a' a'',b'' b'}
\left[\widetilde{\Psi}^{\textrm{BS}}(q)\right]_{a''b''}
\left[S^{\bar{Q}}_{F}(p)\right]_{b'b},
\end{equation}
where $S^{Q}_F$ is the heavy-quark propagator. In our case,
we can take the interaction kernel $K$ to be 
\begin{equation}
\left[K(p-q)\right]_{a' a'',b'' b'}
= \gamma^0_{a'a''} 
[-i\widetilde{V}(\bm{p-q})] \gamma^0_{b''b'},
\label{kernel}
\end{equation}
where 
\begin{equation}
\widetilde{V}(\bm{p-q})
=\int d^3x\, e^{i(\bm{p}-\bm{q})\cdot \bm{x}}V(r)
\end{equation}
is the Fourier transform of the nonrelativistic potential. Then 
we can rewrite the BS equation as
\begin{equation}
\label{BS2}
\widetilde{\Psi}^{\textrm{BS}}_{ab}(p)
=\int \frac{d^4q}{(2\pi)^4}\,
\frac{\left[(/\!\!\! p_Q +m)\gamma^{0}\right]_{aa'}}
      {p_{Q}^2-m^2+i\epsilon}
[i\widetilde{V}(\bm{p-q})]\widetilde{\Psi}^{\textrm{BS}}_{a'b'}(q)
 \frac{\left[\gamma^0(/\!\!\! p_{\bar{Q}} -m)
        \right]_{b'b}
       }
      {p_{\bar{Q}}^2-m^2+i\epsilon}.
\end{equation}

Since we have neglected retardation effects, which are of relative order 
$v^2$, the interaction kernel (\ref{kernel}) is independent of $(p-q)^0$. 
Consequently, the integration over $q^0$ on the right-hand side 
of Eq.~(\ref{BS2}) produces the equal-time BS wave function. Integrating
Eq.~(\ref{BS2}) over $p^0$ to form the equal-time wave
function on the left-hand side, we obtain the Salpeter
equation~\cite{Salpeter:1952ib}:
\begin{equation}
\label{S}
\widetilde{\Psi}^{\textrm{S}}_{ab}(\bm{p})
=\int\frac{dp^0}{2\pi}
\int\frac{d^3q}{(2\pi)^3}
   \frac{\left[(/\!\!\! p_Q +m)\gamma^{0}\right]_{aa'}}
      {p_{Q}^2-m^2+i\epsilon}
\left[i\widetilde{V}(\bm{p}-\bm{q})\right]
\widetilde{\Psi}^{\textrm{S}}_{a'b'}(\bm{q})
 \frac{\left[\gamma^0(/\!\!\! p_{\bar{Q}} -m)
        \right]_{b'b}
       }
      {p_{\bar{Q}}^2-m^2+i\epsilon},
\end{equation}
where
\begin{equation}
\widetilde{\Psi}^{\textrm{S}}_{ab}(\bm{p})=\int\frac{dp^0}{2\pi}
\widetilde{\Psi}^{\textrm{BS}}_{ab}(p)
\end{equation}
is the equal-time BS wave function, which is also known as the Salpeter
wave function. 

We can obtain a wave function that is accurate through corrections of
order $v$ by substituting the trial wave function (\ref{trial-wf}) into
right-hand side of the Salpeter equation (\ref{S}). The result is 
\begin{eqnarray}
\label{Vpsi-x}
\widetilde{\Psi}^{\textrm{S}}_{ab}(\bm{p})
&=&
\int\frac{dp^0}{2\pi}
\int\frac{d^3q}{(2\pi)^3}
\left[i\widetilde{V}(\bm{p}-\bm{q})\widetilde{\psi}(\bm{q})\right]
\nonumber\\
&&
\quad
\quad
\times
\,\,\,
\frac{
\left[
   \left(/\!\!\! p_Q +m\right)
   \gamma^0
\Pi_\lambda
   \gamma^0
   \left(/\!\!\! p_{\bar{Q}} -m\right)
\right]_{ab}
     }
     {\left(p_{Q}^{2}      -m^2+i\epsilon\right)
      \left(p_{\bar{Q}}^{2}-m^2+i\epsilon\right)}.
\label{salpeter-mom}
\end{eqnarray}
Now we substitute Eq.~(\ref{pQ}) into Eq.~(\ref{salpeter-mom})
and carry out the the $p^0$ integral on the right-hand side 
using residue theorem. If we close the contour in the lower half-plane,
then we encounter two poles:
\begin{subequations}
\begin{eqnarray}
p^0&=&\sqrt{m^2+\bm{p}^2}-M_{\rm meson}/2-i\epsilon
\approx  \frac{1}{2}
\left( \frac{\bm{p}^2}{m}-\epsilon_{\textrm{B}}\right)-i\epsilon,
\\
p^0&=&\sqrt{m^2+\bm{p}^2}+M_{\rm meson}/2-i\epsilon
\approx 2m-i\epsilon.
\end{eqnarray}
\end{subequations}
The first pole corresponds to the positive-energy pole in the quark 
propagator, and the second pole corresponds to the negative-energy pole 
in the anti-quark propagator.
The contribution from the residue at the second pole, is suppressed as
$v^2\sim {\bm{p}}^2/m^2 \sim |\epsilon_{\textrm{B}}|/m $ relative to the
contribution from the residue at the first pole. 
Evaluating the latter, we obtain
\begin{eqnarray}
\label{salpeter-mom-2}
\widetilde{\Psi}^{\textrm{S}}_{ab}(\bm{p})
&\approx&
\int\frac{d^3q}{(2\pi)^3}
\left[\widetilde{V}(\bm{p}-\bm{q})\widetilde{\psi}(\bm{q})\right]
\nonumber\\
&&
\quad
\quad
\times
\,\,\,
\frac{
\left.\left[
   \left(/\!\!\! p_{Q} +m\right)
   \gamma^0
\Pi_\lambda
   \gamma^0
   \left(/\!\!\! p_{\bar{Q}} -m\right)
\right]_{ab}\right|_{p^0=0}
     }
     {4m^2\left(\epsilon_{\textrm{B}}-\frac{\bm{p}^2}{2\mu}
-i\epsilon\right)},
\end{eqnarray}
where we have neglected corrections of relative order $v^2$ by
evaluating the numerator at $p^0=0$.
Making use of momentum-space Schr\"odinger equation,
\begin{equation}
\label{mom-sch}
\left(\epsilon_{\textrm{B}}-\frac{\bm{p}^2}{2\mu}\right)
\widetilde{\psi}(\bm{p})
=
\int\frac{d^3q}{(2\pi)^3} \widetilde{V}(\bm{p}-\bm{q})
\widetilde{\psi}(\bm{q}),
\end{equation}
we find that
\begin{equation}
\widetilde{\Psi}^{\textrm{S}}_{ab}(\bm{p})
\approx \frac{1}{4m^2}
   [(/\!\!\! p_Q +m)\gamma^{0}]_{aa'}
  (\Pi_\lambda)_{a'b'}
   [\gamma^0(/\!\!\! p_{\bar{Q}} -m)]_{b'b}
  \widetilde{\psi}(\bm{p}).
\end{equation}
It is easy to see, by substituting this result into the right-hand side
of the Salpeter equation (\ref{S}) and repeating the
preceding analysis, that this is a self-consistent solution of the
Salpeter equation through terms of relative order $v$.

Finally, returning to coordinate space, we can write the equal-time
BS wave functions as 
\begin{subequations}
\begin{eqnarray}
\langle\eta_c (P)|\bar{Q}Q(x^0=0)|0\rangle
&\approx&
-\frac{1}{16\sqrt{2}m^3}
\left(/\!\!\! p_Q+m\right)
   \left(/\!\!\!\! P +2m\right)\gamma^5
\left(/\!\!\! p_{\bar{Q}}-m\right)
\psi (\bm{x}),
\\
\langle J/\psi(P,\epsilon)|\bar{Q}Q(x^0=0)|0\rangle
&\approx&
\frac{1}{16\sqrt{2}m^3}
\left(/\!\!\! p_Q+m\right)
   \left(/\!\!\!\! P +2m\right)
/\!\!\! \epsilon
\left(/\!\!\! p_{\bar{Q}}-m\right)
\psi (\bm{x}).
\end{eqnarray}
\end{subequations}
As we have mentioned, $p$ now has only spatial components. The 
interpretation of $p^j$ in coordinate space is $-i \nabla^j$.
Using $P\cdot p=0$ and $P\cdot \epsilon=0$, we can rewrite the 
equal-time wave functions as 
\begin{subequations}
\label{equal-time-wfs}
\begin{eqnarray}
\langle\eta_c (P)|\bar{Q}Q(x^0=0)|0\rangle
&\approx&
\frac{1}{2\sqrt{2}}\left(\gamma^5+\frac{1}{2m}/\!\!\!\! P \gamma^5
+\frac{1}{4m^2}\left[/\!\!\! p\,\,,/\!\!\!\! P\,\right]\gamma^5
\right)\psi(\bm{x}),
\\
\langle J/\psi (P,\epsilon)|\bar{Q}Q(x^0=0)|0\rangle
&\approx&
-\frac{1}{2\sqrt{2}}
\left(
/\!\!\!\epsilon +\frac{1}{2}\left[\gamma^0,/\!\!\!\epsilon\right]
+\frac{1}{2m}\left\{/\!\!\! p,/\!\!\!\epsilon\right\}
+\frac{i}{2m^2}
\epsilon_{\alpha\beta\gamma\rho}
p^\alpha P^\beta \epsilon^\gamma\gamma^{\rho}\gamma^5
\right)
\nonumber\\
&&\times\psi(\bm{x}),
\end{eqnarray}
\end{subequations}
where we have retained only terms up to those linear in $p$, and 
we have used 
\begin{subequations}
\begin{eqnarray}
\left\{\gamma^\alpha ,\sigma^{\mu\nu}\right\}
&=&-2\epsilon^{\alpha\mu\nu\beta}\gamma_\beta \gamma^5,
\\
/\!\!\! p\,\, /\!\!\!\! P \,/\!\!\! \epsilon
+/\!\!\!\! P \,/\!\!\! \epsilon\, /\!\!\! p
&=& 2i\epsilon^{\alpha\beta\gamma\rho}p_\alpha P_\beta
\epsilon_\gamma\gamma_{\rho}\gamma^5.
\end{eqnarray}
\end{subequations}

By following a procedure similar to the one that we have just 
presented, one can also obtain the BS wave functions at equal light-cone 
time ($x^+=0$). In this case, one integrates the Fourier transform of 
the BS equation (\ref{BS2}) over $p^-$, rather than $p^0$. The 
contribution at leading order in $v$ still comes from the residue of the 
positive-energy pole of the heavy-quark propagator. Now, however, the 
mass-shell condition
\begin{subequations}
\label{mass-shell}
\begin{equation}
\left(P/2+p\right)^0=\sqrt{m^2+\bm{p}^2}
\end{equation}
or, equivalently,
\begin{equation}                                                        
\left(P/2+p\right)^-\left(P/2+p\right)^+
=m^2+\bm{p_\perp}^2
\end{equation}
\end{subequations}
is used to eliminate $p^-$, rather than $p^0$ in the residue. Hence, one
can obtain the equal-light-cone-time wave functions from the equal-time
wave functions by using Eq.~(\ref{mass-shell}) to express $p^3$ in terms
of $p^+$. We work out the details of this transformation in
Sec.~\ref{sec:light-cone}.

Comparing the definitions of the light-cone distribution amplitudes
in Eqs.~(\ref{bs-lc}), (\ref{bs-lc2}), and (\ref{bs-lc3}) with the
equal-time BS wave functions in Eqs.~(\ref{equal-time-wfs}), we obtain
the correspondences 
\begin{subequations}
\label{corr1}
\begin{eqnarray}
       P_A&\leftrightarrow&\psi(r),\\
       P_P&\leftrightarrow&\psi(r),\\
x^- P_T&\leftrightarrow& p^-\psi(r)\approx -p^+\psi(r),\label{P_T}
\end{eqnarray}
\end{subequations}
and 
\begin{subequations}
\label{corr2}
\begin{eqnarray}
V_\bot
&\leftrightarrow&\psi(r),
\\
\widetilde{V} &\leftrightarrow & 0,\label{V-tilde-corres}
\\
V_T&\leftrightarrow&\psi(r),
\\
x^- V_A &\leftrightarrow& p^-\psi(r)\approx -p^+\psi(r).\label{V_A}
\end{eqnarray}
Using Eqs.~(\ref{V-tilde}) and (\ref{V-tilde-corres}), we obtain
\begin{eqnarray}
V_L&\leftrightarrow& \psi(r).
\label{corr4}
\end{eqnarray}
\end{subequations}
The meaning of these correspondences is that we take the Fourier
transform of the quantities on the right-hand sides, use the mass-shell
condition (\ref{mass-shell}) to eliminate $p^3$ in favor of
$z=(P^+/2+p^+)/P^+$, and integrate over $p_\perp$ to form the light-cone
distribution at zero transverse separation. We will give the details of
these steps in Sec.~\ref{sec:light-cone}.

\section{light-cone distribution amplitudes}
\label{sec:light-cone}%
In this section, we compute the light-cone distribution amplitudes in
terms of the potential-model wave functions, making use of the
correspondences in Eqs.~(\ref{corr1}) and (\ref{corr2}).

Let us consider first the cases in which a light-cone distribution
amplitude corresponds to $\psi(r)$, as for $P_A$, $P_P$, $V_\bot$,
$V_T$, and $V_L$. We begin by taking the Fourier transform of $\psi(r)$:
\begin{equation}
\widetilde{\psi}(\bm{p})=
\int d^3x e^{i\bm{p}\cdot \bm{x}}\psi(r).
\label{eq:momentum-psi}
\end{equation}
From the definitions of the light-cone distribution amplitudes in
Eqs.~(\ref{bs-lc}), (\ref{bs-lc2}), and (\ref{bs-lc3}), we see that the
light-cone momentum fraction $z$ is related to the momentum variable $p$
as
\begin{equation}
2z-1=\frac{2n\cdot p}{n\cdot P} \quad\Rightarrow\quad 
z=\frac{n\cdot(P/2+p)}{n\cdot P},
\label{eq:z}
\end{equation} 
where $n=(0^+,1^-,\bm{0}_\perp)$ is a light-like vector whose spatial
component is parallel to the three-momentum $\bm{P}$ of the bound state. 
At the level of precision of this calculation, there is some ambiguity in the 
definition of $z$, in that we can discard terms of relative order $v^2$. 
Hence, we can write in the quarkonium rest frame
\begin{equation}
z=\frac{P^+/2+p^+}{P^+}=\frac{P^+/2+p^+}{P^0}\approx
\frac{P^+/2+p^+}{P^0+2p^0}=\frac{E_Q+p^3}{2E_Q},
\label{z-defn}
\end{equation}
where we have made use of the mass-shell condition (\ref{mass-shell}).
The quark energy $E_Q$ is defined as
\begin{equation}
E_Q=\sqrt{|\bm{p}|^2+m^2_c}.
\label{eq}
\end{equation}
We take the last expression in Eq.~(\ref{z-defn}) as our definition of
$z$. This definition is identical to that which is used in BC. It has
the desirable properties that $0\leq z \leq 1$ and that $z
\leftrightarrow 1-z$ under the interchange of the quark and the
antiquark. In this definition, $m_c$ is the pole mass of the constituent
heavy quark. Note that we distinguish $m_c$ from the parameter $m$ that
appears in the bound-state equation for our potential-model wave
function. For numerical calculations in this paper we take $m_c=1.4$~GeV.

According to the definitions of the light-cone distribution amplitudes
in Eqs.~(\ref{bs-lc}), (\ref{bs-lc2}), and (\ref{bs-lc3}), the
light-cone distribution amplitudes are given by the BS wave functions at
zero transverse spatial separation, as well as zero light-cone-time
separation. Therefore, in momentum space, we obtain the light-cone
distribution amplitudes by integrating over the transverse momenta of
the potential-model wave functions. Carrying out this integration and
making the change of variables to the light-cone momentum fraction $z$,
we find that the light-cone distribution amplitude $\phi(z)$ is given by
\begin{equation}
\phi(z)=\frac{1}{\psi(0)}\int\frac{d^2p_\perp}{(2\pi)^3}\,
\frac{\partial p^3}{\partial z}\, \widetilde{\psi}(\bm{p}),
\label{eq:phi-z}
\end{equation} 
where the prefactor is introduced in order to respect the 
conventional normalization condition
\begin{equation}
\int_0^1 dz\,\phi(z)=1.
\label{eq:phi-norm}
\end{equation}
Hence, we have from Eqs.~(\ref{corr1}) and (\ref{corr2}), 
\begin{equation}
P_A(z)=P_P(z)=V_\bot(z)=V_T(z)=\phi(z). 
\label{lc-dists}
\end{equation}

Now we can simplify the expression for the light-cone distribution amplitude.
The variables $p^3$ and $|\bm{p}|$ can be expressed in terms of $z$
and $p_\perp\equiv|\bm{p}_\perp|$ as
\begin{subequations}
\begin{eqnarray}
p^3&=&\frac{z-\frac{1}{2}}{\sqrt{z(1-z)}}\sqrt{p_\perp^2+m^2_c},
\label{eq:k3}
\\
|\bm{p}|&=&\sqrt{
\frac{p_\perp^2 +4m_c^2\left(z-\frac{1}{2}\right)^2}{4z(1-z)}
           }.
\label{eq:k}
\end{eqnarray}
\end{subequations}
The Jacobian in Eq.~(\ref{eq:phi-z}) is obtained from Eq.~(\ref{eq:k3})
and Eq.~(\ref{eq:k}):
\begin{equation}
\frac{\partial p^3}{\partial z}
=\frac{\sqrt{p_\perp^2+m_c^2}}{4\left[z(1-z)\right]^{3/2}}
=\frac{\sqrt{|\bm{p}|^2+m_c^2}}{2z(1-z)}.
\label{jac-exact}
\end{equation} 
The angular integral in Eq.~(\ref{eq:momentum-psi}) is simple to
evaluate. Furthermore, for a fixed light-cone momentum fraction $z$, we
can make use of Eq.~(\ref{eq:k}) to re-express the integral over the
transverse momentum $\bm{p}_\perp$ in Eq.~(\ref{eq:phi-z}) as an
integral over the $|\bm{p}|$:
\begin{equation}
\int_{-\infty}^\infty d^2p_\perp
=8\pi z(1-z)\int_{\sqrt{d(z)}}^\infty d|\bm{p}| \, |\bm{p}|.
\label{dkperp-dk}
\end{equation}
Then, the expression for the light-cone distribution amplitude reduces to
\begin{equation}
\phi(z)=
\frac{2}{\pi\psi(0)}
\int_{\sqrt{d(z)}}^\infty d|\bm{p}|
\sqrt{|\bm{p}|^2+m_c^2}
\int_0^\infty dr \,\psi(r)\, r \sin(|\bm{p}|r).
\label{eq:phi-z-exact}
\end{equation}
The lower bound of the $|\bm{p}|$ integral is a function of $z$:
\begin{equation}
\sqrt{d(z)}=m_c\frac{\left|z-\frac{1}{2}\right|}{\sqrt{z(1-z)}},
\label{eq:d}
\end{equation}
which is the right-hand side of Eq.~(\ref{eq:k}) at $p_\perp=0$.
The $z$ dependence of $\phi(z)$ appears only in the lower bound $\sqrt{d(z)}$
of the $|\bm{p}|$ integral in Eq.~(\ref{eq:phi-z-exact}). The symmetry
$\phi(z)=\phi(1-z)$ is manifest. Furthermore, it is clear that the 
maximum value of $\phi(z)$ occurs at $z=1/2$ and the minimum value 
(zero) occurs at $z=0$ or $z=1$.

Now let us turn to the light-cone distribution amplitudes $P_T$ and
$V_A$, whose correspondences with the equal-time BS wave functions are
given in Eqs.~(\ref{P_T}) and (\ref{V_A}), respectively. In each of
these cases, the Fourier transform replaces the factor $x^-$ on the left
side of the correspondence with a factor $\partial/\partial p^+$.
Consequently, the solution of the resulting differential equation for
$P_T$ or $V_A$ contains an integration over $p^+$. The expressions for
$P_T$ and $V_A$ that follow from these solutions are not normalizable
because this integration and the factor $p^+$ on the right-hand side of each
correspondence lead to an ultraviolet divergence. Such ultraviolet
divergences are characteristic of amplitudes involving corrections of
higher order in $v$. They can be rendered finite by treating them within
an effective field theory, such as NRQCD. Rather than introduce this
complication into the present discussion, we simply drop such
contributions to the cross section. That is, we drop the contributions
that arise from the light-cone distribution amplitudes $V_A$ and $P_T$.
These contributions are suppressed as $v^2$, and, as we shall see, they
have only about a 30\% affect on the cross-section calculation for the
model light-cone distribution amplitudes that are used in BC. Hence,
these contributions do not have a significant effect on the discussion
of the order-of-magnitude discrepancy between the calculation in BC and
the NRQCD factorization calculations.

\subsection{Nonrelativistic approximation}

In the nonrelativistic approximation, one neglects the relative momenta 
of the heavy quarks in comparison with $m_c$. In the light-cone 
formalism, this amounts to taking  
$\bm{p}_\perp= \bm{0}_\perp$ and $z=1/2$. In this approximation, the 
Jacobian in Eq.~(\ref{jac-exact}) reduces to 
\begin{equation}
\left[\frac{\partial p^3}{\partial z}\right]_{\textrm{approx}}
=2m_c.
\label{jac-approx}
\end{equation}
The corresponding approximate 
version of Eq.~(\ref{eq:phi-z-exact}) is 
\begin{equation}
\phi_{\textrm{approx}}(z)=
z(1-z)
\frac{8m_c}{\pi\psi(0)}
\int_0^\infty dr \,\psi(r)\, \cos\left[\sqrt{d(z)}r\right].
\label{eq:phi-z-approx}
\end{equation}
The approximate Jacobian (\ref{jac-approx}) was used in BC. For a given
potential-model wave function, it leads to a narrower light-cone
distribution amplitude than does the exact Jacobian (\ref{jac-exact}).
Since, in this paper, we are generally working at leading order in
$v$, it is unnecessary (but not inconsistent) to include the corrections
of higher order in $v$ that are contained in the exact Jacobian.
However, it is convenient to use the exact Jacobian because it preserves
the normalizations of the light-cone distributions, and we have used it
to derive all of the numerical results that are presented in this paper.
As a check, we have also carried out numerical calculations using the
approximate Jacobian (\ref{jac-approx}), and those calculations are
qualitatively consistent with the conclusions that we draw from the
calculations that are based on the exact Jacobian.

\subsection{Expressions for the light-cone distribution amplitudes for
efficient numerical evaluation}
The double integral in Eq.~(\ref{eq:phi-z-exact}) is difficult to 
evaluate numerically because of the oscillatory nature of the inner 
integrand and because the integrand of the outer integral 
decreases only slowly as $|\bm{p}|\to \infty$. We can improve the numerical 
accuracy of the integrations by splitting the integral into two pieces:
\begin{equation}
\phi(z)=\phi_a(z)+\phi_b(z),
\end{equation}
where 
\begin{subequations}
\label{eq:phiab-z-exact}
\begin{eqnarray}
\phi_a(z)&=&
\frac{2}{\pi\psi(0)}
\int_{\sqrt{d(z)}}^\infty d|\bm{p}|
\,|\bm{p}|
\int_0^\infty dr \,\psi(r)\, r \sin(|\bm{p}|r),
\label{eq:phia-z-exact}
\\
\phi_b(z)&=&
\frac{2}{\pi\psi(0)}
\int_{\sqrt{d(z)}}^\infty d|\bm{p}|
\left(\sqrt{|\bm{p}|^2+m_c^2}-|\bm{p}|\right)
\int_0^\infty dr \,\psi(r)\, r \sin(|\bm{p}|r).
\label{eq:phib-z-exact}
\end{eqnarray}
\end{subequations}
The integrand in $\phi_a(z)$ contains the leading behavior of 
the integrand in Eq.~(\ref{eq:phi-z-exact}), but is a simpler form in 
which the $|\bm{p}|$ integration can be carried out analytically. The integral 
$\phi_b(x)$ contains the remainder, whose integrand converges more rapidly as 
$|\bm{p}|\to\infty$. We carry out the integration over $|\bm{p}|$ in 
$\phi_a(z)$, treating the divergent integration as a Fourier transform whose 
result is defined in the distribution sense. The result is 
\begin{equation}
\phi_a\left(z\right)
=
1+\frac{2}{\pi\psi(0)}
\int_0^\infty dr\,\psi(r)
\left\{
\sqrt{d(z)}\cos \left[\sqrt{d(z)} r\right]
-\frac{\sin\left[\sqrt{d(z)} r\right]}{r}
\right\}.
\label{eq:phia-z-single}
\end{equation}
We use Eqs.~(\ref{eq:phib-z-exact}) and (\ref{eq:phia-z-single}) to 
evaluate $\phi(z)$ numerically. 

It is useful, in checking the accuracy of numerical-integration methods
that we use, to know the values of the light-cone distribution amplitudes for
special cases in which they can be computed analytically. The light-cone
distribution amplitudes $\phi$, $\phi_a$, and $\phi_b$ are computed for the 
case of a pure Coulomb potential in Appendix~\ref{App:Coulomb}. Their
asymptotic behaviors at $z=0$ (or $z=1$) are also given. In addition, it
is simple to compute the value of $\phi_a$ at $z=1/2$. In this case
$d(z)=0$, and Eq.~(\ref{eq:phia-z-single}) reduces to 
\begin{equation}
\phi_a(1/2)=1.
\end{equation}

\subsection{Light-cone distribution amplitudes}

\begin{figure}[t]
\includegraphics[width=12cm]{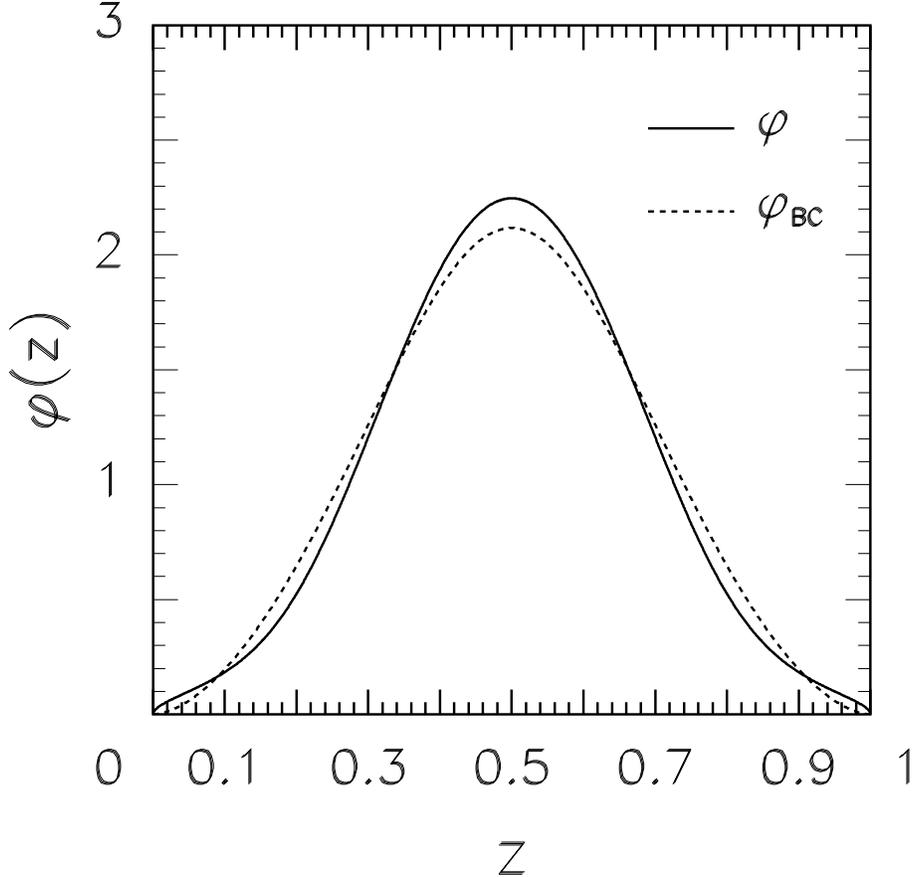}
\caption{The light-cone distribution amplitude $\phi(z)$ that is derived 
from the 
potential model in this paper and the model light-cone distribution amplitude
$\phi_{\rm BC}(z)$ that is used in BC. As is explained in the text, 
$\phi(z)$ is computed from the potential-model wave
function by carrying out the integrations in 
Eqs.~(\ref{eq:phib-z-exact}) and (\ref{eq:phia-z-single}) 
numerically, taking $m_c=1.4$~GeV. 
$\phi_{\rm BC}(z)$ is given in Eq.~(\ref{phi-bc}).}
\label{fig:BC-Pot-dists}
\end{figure}

In Fig.~\ref{fig:BC-Pot-dists}, we show the light-cone distribution amplitude 
that results from substituting our potential-model wave
function into Eqs.~(\ref{eq:phib-z-exact}) and (\ref{eq:phia-z-single}) 
and carrying out the integrations numerically, taking $m_c=1.4$~GeV.
For comparison, we also show the model light-cone distribution amplitude 
that is used in BC:
\begin{equation}
\phi_{\rm BC}(z)
=c(v^2)z(1-z)\left[\frac{z(1-z)}{1-4z(1-z)(1-v^2)}\right]^{1-v^2},
\label{phi-bc}
\end{equation}
where $v^2=0.3$ and $c(0.3)\approx 9.62$.
As can be seen, the model light-cone distribution amplitude of BC is
very similar in shape to the light-cone distribution amplitude that we have
derived from a potential model. However, there is a significant
difference in the functional forms in the tails at $z=0,1$. The
asymptotic behavior of the BC light-cone distribution amplitude at $z=0$ is
$\phi_{\rm BC}(z)\sim c(v^2)z^{(2-v^2)}$. In contrast,
the potential-model light-cone distribution amplitude that we derive 
behaves at $z=0$ as $\phi(z)\sim 8\gamma_C z^{1/2}/(\pi m_c)$
[Eq.~(\ref{asymptotics})].

\begin{figure}[t]
\includegraphics[width=12cm]{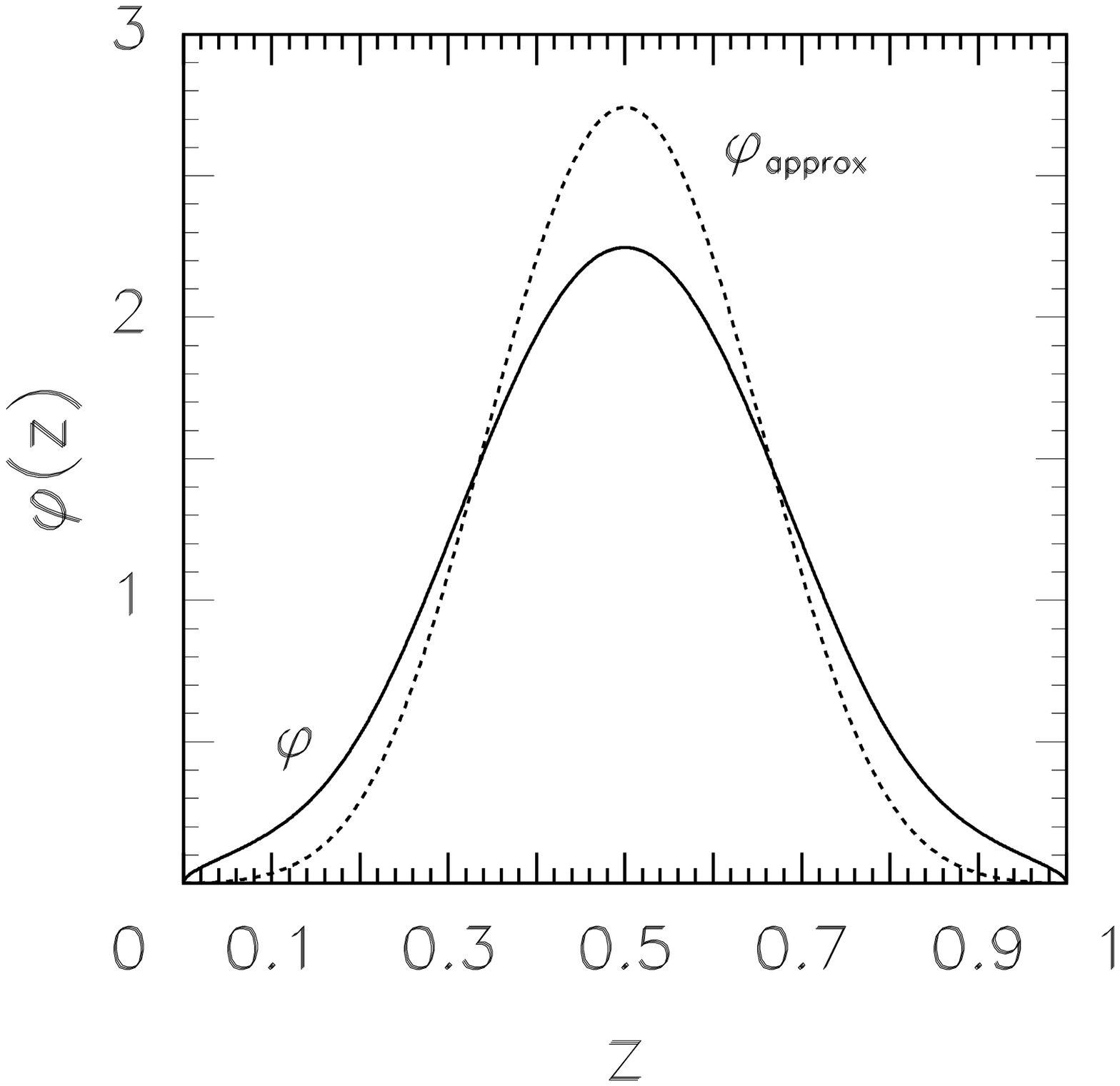}
\caption{Comparison of the light-cone distribution amplitude 
$\phi_{\rm approx}(z)$ with $\phi(z)$. $\phi_{\rm approx}(z)$ is computed from
Eq.~(\ref{eq:phi-z-approx}), in which the nonrelativistic approximation
is used for the light-cone-fraction Jacobian.  $\phi(z)$ is
computed from the exact expressions in Eqs.~(\ref{eq:phib-z-exact}) and
(\ref{eq:phia-z-single}).}
\label{fig:phi-phiapprox}
\end{figure}

In Fig.~\ref{fig:phi-phiapprox}, we show the light-cone distribution amplitude
$\phi_{\rm approx}(z)$ that arises from Eq.~(\ref{eq:phi-z-approx}), in
which the nonrelativistic approximation has been used for the
light-cone-fraction Jacobian.
As can be seen from the figure, this approximation leads to a narrower
light-cone distribution than does the exact expression.

\section{Perturbative correction contained in the light-cone
distribution amplitude}
\label{sec:pert-corr}%
As can be seen from Eq.~(\ref{eq:k}),
when $\phi(z)$ is used in computing the cross section, it yields 
contributions that arise from both large and small $Q\bar Q$ 
relative momentum $p$. The large-$p$ contributions are not treated 
correctly in a potential model, which is inherently nonrelativistic. 
In the NRQCD factorization approach, this difficulty is dealt with 
by factoring the large-$p$ contributions into short-distance 
coefficients and the small-$p$ contributions into NRQCD matrix 
elements. The large-$p$ contributions are then computed in 
perturbation theory, which is presumed to be valid, owing to 
asymptotic freedom. 

In the light-cone calculation, we also wish to identify, and ultimately
remove, these large-$p$ contributions. In particular, we wish to remove
that part of the large-$p$ contribution that, in the NRQCD approach, is
computed as an order-$\alpha_s$ contribution to the cross section, {\it
i.e.}, an order-$\alpha_s$ contribution to the short-distance
coefficient. Of course, we could, in principle, remove contributions 
of order $\alpha_s^2$ and higher, as well. These contributions, 
presumably, are less important than the order-$\alpha_s$ contribution.

\begin{figure}
\includegraphics[width=5.5cm]{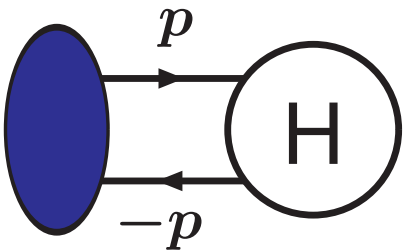}
\caption{Diagrammatic representation of $\mathcal{M}$, as given in 
Eq.~(\ref{H-and-wavefn}). 
The circle labeled $H$ represents the hard part of the production
amplitude, and the oval represents the quarkonium wave function.
\label{fig:H-and-wavefn}}
\end{figure}

\begin{figure}[t]
\includegraphics[width=\textwidth]{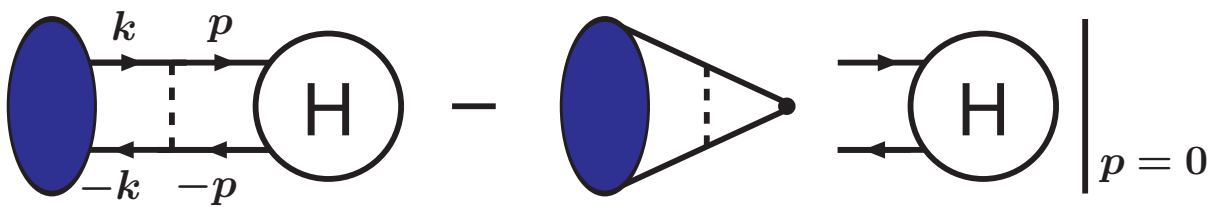}
\caption{Diagrammatic representation of $\mathcal{M}-\mathcal{M}_0$, 
as given in Eq.~(\ref{subtraction}). 
The circle labeled $H$ represents the hard part of the production
amplitude, and the oval represents the quarkonium wave function.
\label{fig:subtraction}}
\end{figure}

We begin by writing the production amplitude for the
heavy-quarkonium in a schematic form, which is depicted in 
Fig.~\ref{fig:H-and-wavefn}:
\begin{equation}
\mathcal{M} = \int\frac{d^4p}{(2\pi)^4}2\pi\delta(p^0)
\widetilde{\psi}(\bm{p}) H(p),
\label{H-and-wavefn}
\end{equation}
where $\widetilde{\psi}(\bm{p})$ is the spatial part of the
momentum-space wave function for the meson and $H(p)$ is the
short-distance production amplitude for the quark pair. 
Here we approximate the temporal part of the wave function with
$2\pi\delta(p^0)$. That is, we neglect deviations of the quark energy
from $m$, which are of relative order $v^2$. In the rest frame of the
meson, $\bm{p}$ is the momentum of the charm quark and $-\bm{p}$ is that
for the antiquark. At this point, we could simply impose a cutoff on the
integration over $\bm{p}$ in order to separate the large- and small-$p$
contributions. However, we wish to maintain consistency with dimensional
regularization, which is conventionally used to compute the
short-distance coefficients in perturbation theory in the NRQCD method.
Therefore, we use the method of regions \cite{Beneke:1997zp} to compute
the large and small-$p$ contributions. For small $p$, we approximate
$H(p)$ by $H(0)$. Then, we obtain the standard leading-order
contribution to the cross section:
\begin{equation}
\mathcal{M}_0 = \int\frac{d^3p}{(2\pi)^3}
\widetilde{\psi}(\bm{p}) H(0)
= \psi(0) H(0).
\end{equation}
We subtract $\mathcal{M}_0$ from $\mathcal{M}$. This removes a
scaleless power infrared divergence, which vanishes in dimensional
regularization. The remainder is infrared finite, and it is dominated
by large $p$. Therefore, we can use the Bethe-Salpeter equation with
the Coulomb-exchange kernel to write $\widetilde{\psi}(\bm{p})$
approximately in a form in which one Coulomb loop is exposed. The
result, which is depicted in Fig.~\ref{fig:subtraction}, is
\begin{eqnarray}
\mathcal{M}-\mathcal{M}_0
&=&\int\frac{d^4k}{(2\pi)^4}2\pi\delta(k^0)\widetilde{\psi}(\bm{k})
\int\frac{dp^0d^3p}{(2\pi)^4}
\left(-ig_sT^a\right)
\left(+ig_sT^a\right)
\,
\frac{i}{(\bm{p}-\bm{k})^2}
\,
\frac{i}{p^0-\frac{\bm{p}^2}{2m}+i\epsilon}
\,\nonumber\\
&&\qquad\times\frac{-i}{p^0+\frac{\bm{p}^2}{2m}-i\epsilon}
[H(p)-H(0)],
\label{subtraction}
\end{eqnarray}
where $g_s$ is the strong coupling, and $T^a$ is an SU(3) color matrix.
Note that we use the nonrelativistic expressions for the $Q$ and $\bar
Q$ propagators because we wish to identify the contribution to the
high-momentum tail of the light-cone distribution that arises in the
potential model. In the potential model, the heavy quark and antiquark
are treated in the leading nonrelativistic approximation. Taking the
small-$k$ approximation for the integrand of the $\bm{k}$ integration, 
we obtain the order-$\alpha_s$ short-distance contribution to 
$\cal M$:
\begin{equation}
\Delta\mathcal{M}
=\psi(0)\int\frac{dp^0d^3p}{(2\pi)^4}
\left(-ig_sT^a\right)
\left(+ig_sT^a\right)
\,
\frac{i}{\bm{p}^2-i\epsilon}
\,
\frac{i}{p^0-\frac{\bm{p}^2}{2m}+i\epsilon}
\,
\frac{-i}{p^0+\frac{\bm{p}^2}{2m}-i\epsilon}
[H(p)-H(0)].
\label{delta-M-unint}
\end{equation}

We emphasize that $\Delta\mathcal{M}$ is not arbitrary, but has a
precise definition in NRQCD: $\Delta\mathcal{M}/\psi(0)$ is the
order-$\alpha_s$ short-distance coefficient of the NRQCD operator
$\psi\chi^\dagger$ in a QCD-like theory in which the $Q$ and $\bar Q$
are treated in the leading nonrelativistic approximations and the only
interactions are Coulomb-gluon ladder exchanges. That theory is,
of course, identical to the potential model.
$\Delta\mathcal{M}/\psi(0)$ is equal to the diagrams in
Fig.~\ref{fig:subtraction}, with the wave function truncated and $k$ set
to zero. That equality is precisely the NRQCD matching condition: The
first diagram in Fig.~\ref{fig:subtraction} is the order-$\alpha_s$
correction to the on-shell hard-scattering in the full theory, and the
second diagram in Fig.~\ref{fig:subtraction} is the order-$\alpha_s$
correction to the NRQCD operator times the leading-order short-distance
coefficient. (We assume that the on-shell matrix element of
$\psi\chi^\dagger$, which multiplies $\Delta\mathcal{M}/\psi(0)$, is
normalized to unity.)

The integral over $p^0$ in Eq.~(\ref{delta-M-unint}) can be carried
out by the contour method. Here, we assume that any singularities in
$H(p)$ are of order $m$ away from the origin and, therefore correspond to
contributions that are suppressed as $\bm{p}^2/m^2\sim v^2$ relative to
the leading contributions, which come from the poles at
$p^0=\pm\bm{p}^2/(2m)$. The result is then
\begin{subequations}
\begin{eqnarray}
\Delta\mathcal{M}
&=&
4\pi\alpha_s C_F \psi(0)
\int\frac{d^3p}{(2\pi)^3}\frac{m}{\bm{p}^4} 
\left[H(\bm{p})-H(\bm{0})\right]
\nonumber\\
&=&
4\pi\alpha_s C_F \psi(0)
\int\frac{d^3p}{(2\pi)^3}\frac{m}{\left[\bm{p}^4\right]_{+\bm{p}}} 
H(\bm{p}),
\label{delta-M}
\end{eqnarray}
where 
\begin{equation}
H(\bm{p})=H(p)|_{p^0=0}.
\end{equation}
\end{subequations}
Here, we have neglected deviations of $p^0$ from $0$ in $H(p)$, with
errors of relative order $v^2$. The second equality in
Eq.~(\ref{delta-M}), defines the $+$ distribution with respect to
$\bm{p}$. We use the $+$ distribution to write $\Delta \mathcal{M}$ as a
convolution of the correction to the momentum-space wave function
$\Delta\widetilde{\psi}({\bm{p}})$ 
with the hard-scattering amplitude $H(\bm{p})$:
\begin{subequations}
\begin{equation}
\Delta \mathcal{M}=
\int\frac{d^3p}{(2\pi)^3} 
\Delta\widetilde{\psi}(\bm{p})
H(\bm{p}),
\end{equation}
where
\begin{equation}
\Delta\widetilde{\psi}(\bm{p})
=
4\pi\alpha_s C_F \psi(0)
\frac{m}{[\bm{p}^4]_{+\bm{p}}}.
\end{equation}
\end{subequations}
Therefore, we can remove contributions to the production amplitude that
correspond to the standard order-$\alpha_s$ corrections to the NRQCD
short-distance coefficient by modifying the momentum-space wave
function according to
$\widetilde{\psi}(\bm{p})\to 
\widetilde{\psi}(\bm{p})-\Delta\widetilde{\psi}(\bm{p})$.

Similarly, we modify the light-cone distribution amplitude $\phi(z)$ 
according to $\phi(z)\to \phi(z)-\Delta\phi(z)$. The definition of 
$\Delta\phi(z)$ follows from the definition of $\phi(z)$ in 
Eq.~(\ref{eq:phi-z}):
\begin{eqnarray}
\Delta\phi(z)&=&\frac{1}{\psi(0)}\int\frac{d^2p_\perp}{(2\pi)^3}\,
\frac{\partial p^3}{\partial z}\, \Delta\widetilde{\psi}(\bm{p})
\nonumber\\
&=&
4\pi m\alpha_s C_F
\int\frac{d^2p_\perp}{(2\pi)^3}\,
\left[\frac{\partial p^3}{\partial z}\, \frac{1}{\bm{p}^4}\right]_{+z},
\label{dphi-def}
\end{eqnarray} 
The $+$ distribution with respect to $z$ in the last line of 
Eq.~(\ref{dphi-def}) is defined by
\begin{equation}
\int_0^1 dz\,[f(z)]_{+z}H(z)=\int_0^1 dz\,f(z)[H(z)-H(1/2)].
\end{equation}
The evaluation of $\Delta\phi(z)$ can be carried out in analogy with 
the derivation of Eq.~(\ref{eq:phi-z-exact}) 
from Eq.~(\ref{eq:phi-z}). The exact Jacobian in Eq.~(\ref{jac-exact})
is substituted into Eq.~(\ref{dphi-def}) and the $\bm{p}_\perp$ integral
can be written as an integral over $|\bm{p}|$, as in Eq.~(\ref{dkperp-dk}).
Then, the expression for $\Delta\phi(z)$ can be written as a single 
integral:
\begin{equation}
\Delta\phi(z)=\frac{2m\alpha_s C_F}{\pi}
\int_{\sqrt{d(z)}}^\infty 
d|\bm{p}|\left[\frac{\sqrt{|\bm{p}|^2+m_c^2}}{p^3}\right]_{+z}.
\label{dphi-p}
\end{equation}
Using the identity
\begin{equation}
\int \frac{\sqrt{x^2+1}}{x^3} dx=
-
\left[
\frac{\sqrt{x^2+1}}{2x^2}
+ \frac{1}{4} \log\frac{\sqrt{x^2+1}+1}{\sqrt{x^2+1}-1}
\right],
\end{equation}
we obtain an analytic expression for $\Delta\phi(z)$:
\begin{equation}
\Delta\phi(z)=
\frac{2\alpha_s C_F}{\pi}
\left(\frac{m}{m_c}\right)
\left\{
\left[\frac{\sqrt{z(1-z)}}
     {4(z-\frac{1}{2})^2}\right]_{+z}
+\frac{1}{4}
\left[\log\frac{1+2\sqrt{z(1-z)}}{1-2\sqrt{z(1-z)}}
\right]_{+z}
\right\}.
\label{dphi-exact}
\end{equation}
We note that, because of the definition of $\Delta\phi$ as a $+$ 
distribution, $\Delta\phi$ has the property
\begin{equation}
\int_0^1 dz\, \Delta\phi(z)=0.
\end{equation}
That is, the effect of subtracting $\Delta\phi$ from $\phi$ is to shift
contributions from $z$ near $0$ or $1$ to $z$ near $1/2$ without
changing the overall normalization of the light-cone distribution
amplitude. 

As can be seen from Eq.~(\ref{dphi-exact}), $\Delta\phi$ diverges near
$z=1/2$ as $1/(z-1/2)^2$. When $\Delta\phi$ is multiplied by a function
$f(z)$ and integrated over $z$, the leading linear divergence at $z=1/2$
is canceled by the factor $f(z)-f(1/2)$ that appears in the integrand by
virtue of the definition of the $+$ distribution. The subleading
logarithmic divergence cancels because the corresponding integrand
is odd about the point $z=1/2$.

Since the Coulomb potential controls the behavior of the wave function
at large momenta (short distances), we expect the asymptotic behavior of
$\Delta\phi(z)$ near $z=0$ (or $z=1$) to be identical to that of
$\phi(z)$. It is easy to see from Eq.~(\ref{dphi-exact}), that the
asymptotic behavior of $\Delta\phi(z)$ near $z=0$ is
\begin{equation}
\Delta\phi(z)\sim
\frac{8\gamma_{\textrm{C}}}{\pi m_c}\sqrt{z},
\label{dphi-01}
\end{equation}
where we have made use of Eq.~(\ref{gamma-C}). (The asymptotic behavior
near $z=1$ can be obtained from Eq.~(\ref{dphi-01}) by replacing $z$ 
with $1-z$.) As expected, the asymptotic behavior of $\Delta\phi(z)$ 
in Eq.~(\ref{dphi-01}) is identical to that of $\phi(z)$ in 
Eq.~(\ref{asympt-phi}).

\begin{figure}[t]
\includegraphics[width=12cm]{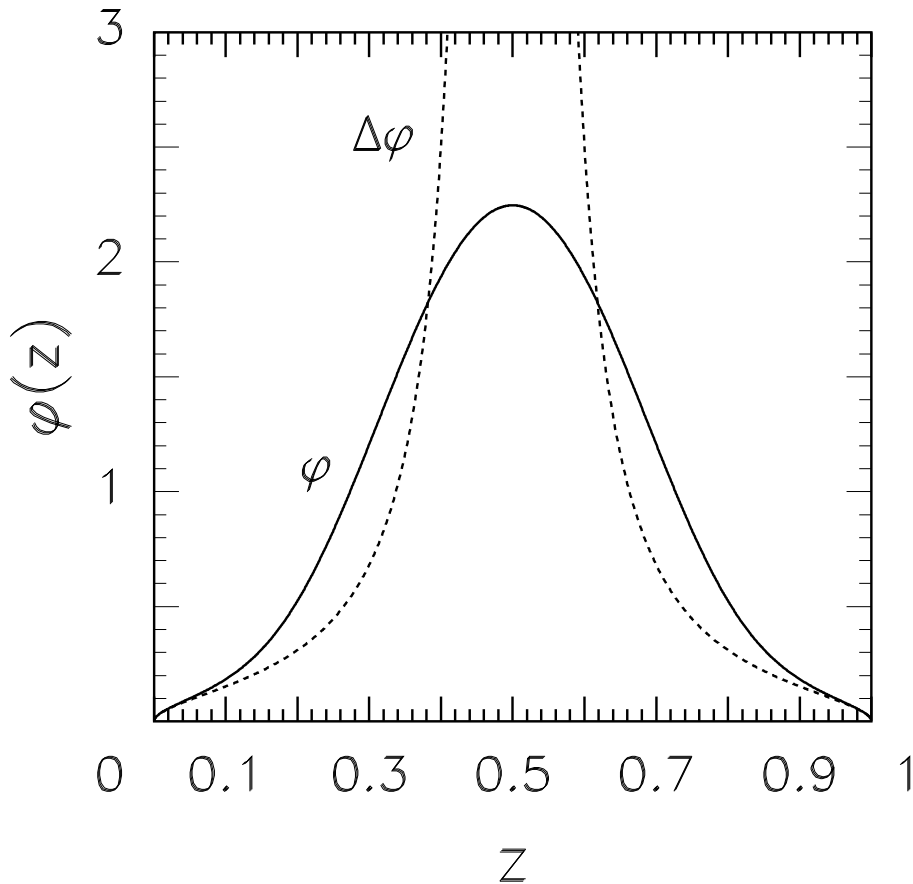}
\caption{$\Delta\phi(z)$ and $\phi(z)$, plotted over the full range of
$z$. $\Delta\phi(z)$ is given by Eq.~(\ref{dphi-exact}) and $\phi(z)$ is
computed from the potential-model wave function by carrying out the
integrations in Eqs.~(\ref{eq:phib-z-exact}) and
(\ref{eq:phia-z-single}) numerically. Note that $\Delta\phi$ is actually
a $+$ distribution, as is described in the text, and, therefore, it
contains a large negative contribution at $z=1/2$ that is not shown in
the figure.}
\label{fig:dphi-full}
\end{figure}

\begin{figure}[t]
\includegraphics[width=12cm]{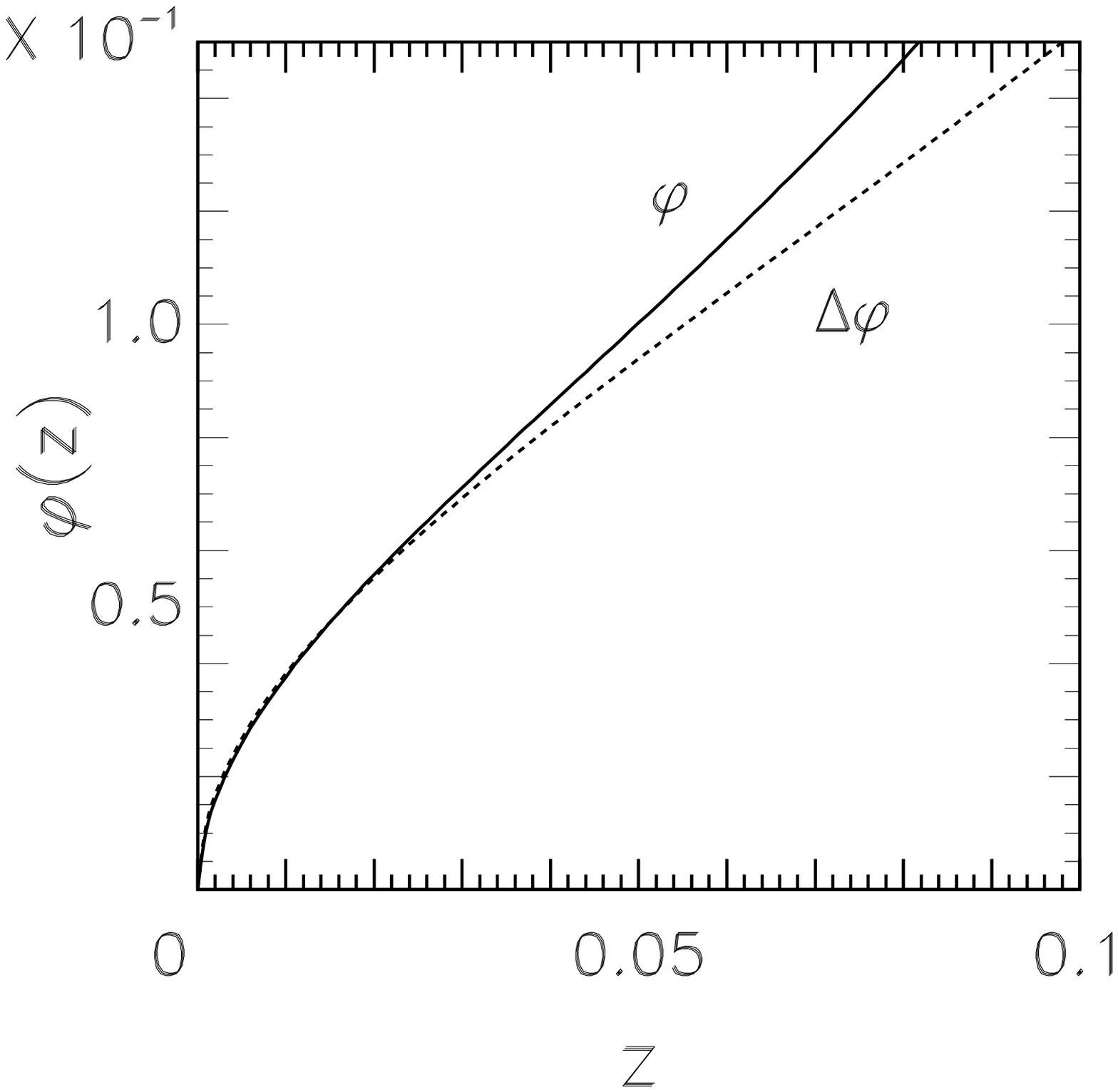}
\caption{$\Delta\phi(z)$ and $\phi(z)$, plotted over a restricted range
in $z$. $\Delta\phi(z)$ is given by Eq.~(\ref{dphi-exact}) and $\phi(z)$
is computed from the potential-model wave function by carrying out the
integrations in Eqs.~(\ref{eq:phib-z-exact}) and
(\ref{eq:phia-z-single}) numerically.}
\label{fig:dphi-small}
\end{figure}

In Figs.~\ref{fig:dphi-full} and \ref{fig:dphi-small}, we plot $\Delta\phi(z)$, 
along with $\phi(z)$. 
Since $\Delta\phi$ is a $+$ distribution [Eq.~(\ref{dphi-exact})], it
contains a large negative contribution at $z=1/2$ that is not shown in
Fig.~\ref{fig:dphi-full}.  As can be seen from Figs.~\ref{fig:dphi-full} and
\ref{fig:dphi-small}, $\Delta\phi(z)$ shows the expected approach to $\phi(z)$
at $z=0$ and $z=1$. In Fig.~\ref{fig:dphi-small}, we show this behavior in
more detail near $z=0$. The convergence of $\phi(z)$ and $\Delta\phi(z)$
at the end points has a significant consequence in the evaluation of the
cross section. As we have mentioned, we must subtract $\Delta\phi(z)$
from $\phi(z)$ in order to remove contributions that are conventionally
calculated as part of the order-$\alpha_s$ contributions to the NRQCD
short-distance coefficient. As we shall see, the cancellation between
$\phi(z)$ and $\Delta\phi(z)$ at the end points results in great
reduction in the cross section.
\section{Light-cone cross section for $\bm{e^+e^-\to J/\psi+\eta_c}$ at
$\bm{B}$ factories}
\label{sec:cross-sections}%
In this section, we use the light-cone distribution amplitude that we have 
derived from the potential model to compute the cross section for 
$e^+e^-\to J/\psi+\eta_c$ at the $B$ factories.
\subsection{Light-cone cross section}
We begin by recording the expression for the light-cone cross section
from BC. In order to compare with the results of BC, we have tried to
make use of the same approximations that were employed in that paper.
(In some cases, it was necessary to deduce the approximations that were
taken in BC by exploring the effects of various approximations on the
numerical results.) In this section, we make use of the notation of BC.
In particular, we use $\bar{M}_Q$ to 
represent the heavy-quark mass, and we distinguish its numerical value 
from those of the parameter $m$ in the potential model and the 
charm-quark pole mass $m_c$.

The light-cone cross section is derived by standard techniques 
\cite{Brodsky:1981kj,Chernyak:1983ej} and is given in BC as
\begin{equation}
\sigma(e^+e^-\to J/\psi+\eta_c)
=\frac{\pi\alpha^2 e_c^2}{6}
\left(
\frac{|\bm{p}|}{E_{\textrm{beam}}}
\right)^3
|F_{VP}(s)|^2,
\label{light-cone-x-section}
\end{equation}
where $\bm{p}$ is the three momentum of the $J/\psi$ in the 
$e^+e^-$ center-of-momentum (CM) frame and 
$\sqrt{s}=2E_{\textrm{beam}}$ is
the $e^+e^-$ CM energy.  The form factor $F_{VP}(s)$
is given by
\begin{equation}
|F_{VP}(s)|=\frac{32\pi}{9}
\left|\frac{f_V f_P\bar{M}}{q_0^4}\right|\,I_0,
\label{sigma}
\end{equation}
where $\bar{M}=(3M_{J/\psi}+M_{\eta_c})/4$ is spin-averaged $S-$wave
charmonium mass and $f_V$ and $f_P$ are the decay constants of 
the $J/\psi$ and the $\eta_c$, respectively. The quantities $|\bm{p}|$ 
and $q_0^2$ are given by
\begin{subequations}
\begin{eqnarray}
|\bm{p}|&=&\sqrt{E_{\textrm{beam}}^2-\bar{M}^2},
\\
q_0^2&=&(E_{\textrm{beam}}+|\bm{p}|)^2.
\end{eqnarray}
\end{subequations}
$I_0$ in Eq.~(\ref{sigma}) is defined by
\begin{equation}
I_0=\int^1_0 dx_1 \int^1_0 dy_1\alpha_s(k^2) \sum_{i=1}^5 f_i(x,y),
\label{I0}
\end{equation}
where $k$ is the momentum of the virtual gluon. The labeling of the
light-cone momentum fractions for the $J/\psi$, $x_1$, $x_2=1-x_1$, and
for the $\eta_c$, $y_1$, $y_2=1-y_1$, can be read from Fig.~1 of BC. The
integrands $f_i(x,y)$ in Eq.~(\ref{I0}) are given by
\begin{subequations}
\label{fi}
\begin{eqnarray}
f_1(x,y)&=&
\frac{Z_t Z_p V_{T}(x) P_{P}(y)}{d(x,y)\, s(x)},
\label{f1}
\\
f_2(x,y)&=&-\frac{\bar{M}_Q^2}{{\bar{M}}^2}\,
\frac{Z_m^\sigma Z_t V_T(x) P_A(y)}{d(x,y)\,s(x)},
\label{f2}
\\
f_3(x,y)&=&
\frac{1}{2}\frac{V_{L}(x)\,P_{A}(y)}{d(x,y)},
\label{f3}
\\
f_4(x,y)&=&\frac{1}{2}
\frac{(1-2y_1)}{s(y)}\frac{V_{\perp}(x)\,P_{A}(y)}{d(x,y)},
\label{f4}
\\
f_5(x,y)&=&
\frac{1}{8} \left( 1-Z_t Z_m^k\frac{4{\bar{M}}_Q^2}{{\bar{M}}^2 }
            \right)\,
\frac{(1+y_1)V_A(x)P_A(y)}{d^2(x,y)}.
\label{f5}
\end{eqnarray}
\end{subequations}
The function $d(x,y)$ in Eq.~(\ref{fi}) originates from 
the gluon propagator in Fig.~1 of BC and is defined by
\begin{subequations}
\label{d}
\begin{eqnarray}
d(x,y)&=&\frac{k^2}{q_0^2}=\left( x_1+\frac{\delta}{y_1}\biggr )
\biggl (y_1+\frac{\delta}{x_1}\right),
\\
\delta&=&\left(Z_m^k \frac{{\bar{M}}_Q}{q_0}\right)^2.
\end{eqnarray}
\end{subequations}
The functions $s(x)$ and $s(y)$ in Eq.~(\ref{fi}) are from 
the charm-quark propagator in Fig.~1 of BC and are defined 
by
\begin{subequations}
\label{s}
\begin{eqnarray}
s(x)&=& x_1+\frac{Z_m^{\sigma}{\bar{M}}_Q^2}{y_1y_2\,q_0^2},
\\
s(y)&=& y_1+\frac{Z_m^{\sigma}{\bar{M}}_Q^2}{x_1x_2\,q_0^2}.
\end{eqnarray}
\end{subequations}
One might suppose that these expressions should contain two factors of
$Z_m^k$ for each factor of ${\bar{M}}_Q^2$, as is the case in the
corresponding expressions in Refs.~\cite{Braguta:2005kr,Braguta:2006nf}.
Here, we wish to compare directly with the results of BC, and so we use
the expressions that are given in that paper. The inclusion of two
factors of $Z_m^k$ for each factor of ${\bar{M}}_Q^2$ in Eq.~(\ref{s}) has
a small numerical effect on the cross section, which we will describe in
Sec.~\ref{sec:discussion}. Following BC, we also take
\begin{subequations}
\begin{eqnarray}
E_{\textrm{beam}}&=&\frac{\sqrt{s}}{2},
\\
|\bm{p}|&\approx&E_{\textrm{beam}}
-\frac{\bar{M}^2}{2E_{\textrm{beam}}},
\\
q_0&\approx&\sqrt{s-2\bar{M}^2},
\\
k^2&\approx& x_1 y_1 q_0^2+2\bar{M}_Q^2.
\end{eqnarray}
\end{subequations}

The various renormalization factors that appear in Eqs.~(\ref{fi}) 
and (\ref{s}) are defined by
\begin{subequations}
\label{Zi}
\begin{eqnarray}
\quad Z_p&=&\left[\frac{\alpha_s(k^2)}{\alpha_s(\bar{M}_Q^2)} 
          \right]^{-3C_F/b_0},
\\
 Z_{t}&=&\left[\frac{\alpha_s(k^2)}{\alpha_s(\bar{M}_Q^2)} 
         \right]^{C_F/b_0},
\\
 Z_{m}(\mu^2)&=&\left[\frac{\alpha_s(\mu^2)}{\alpha_s({\bar{M}}_Q^2)} 
                \right]^{3C_F/b_0},
\\
Z_m^k&=&Z_m(\mu^2=k^2),
\\
Z_m^{\sigma}&=&Z_m(\mu^2=\sigma^2),
\end{eqnarray}
\end{subequations}
where $b_0=25/3$. The running $\overline{\textrm{MS}}$ mass in 
Eqs.~(\ref{Zi}) is given by
\begin{equation}
{\bar{M}}_Q(\mu^2)=Z_m(\mu^2)\,{\bar{M}}_Q.
\end{equation}
Following BC, we take the strong coupling $\alpha_s$ 
to be given approximately by 
\begin{equation}
\alpha_s(\mu^2)\approx
\frac{4\pi/b_0}{\log\left[\mu^2/(200\textrm{MeV})^2\right]}.
\end{equation}
In Eqs.~(\ref{Zi}), $\sigma^2$ is the square of the four-momentum of the 
charm-quark propagator in Fig.~1 of BC. Again, we follow BC and use the 
approximate expression 
\begin{equation}
\sigma^2 \approx 
xq_0^2+{\bar{M}}_Q^2
\left[1+\frac{1}{y(1-y)}\right].
\end{equation}

The renormalization factors $Z_i$ in Eq.~(\ref{Zi}) were introduced 
in BC in order to account for the evolution of the heavy-quark mass 
and tensor and pseudoscalar currents from the scale $\bar{M}_Q$ to 
the scales of the hard interactions in the production process. 
We note that such renormalization factors do not appear 
in the NRQCD calculations of the
cross section. As we shall see, they have a large numerical effect on
the cross section.

We use the following numerical values, which were given in BC:
$s=112$~GeV$^2$, $f_V=f_P=400$~MeV, $\bar{M}_Q=\bar{M}_c=1.2$~GeV. We
also take $M_{J/\psi}=3.096916$~GeV and $M_{\eta_c}=2.98040$~GeV, which
imply that $\bar{M}=3.06779$~GeV. We note that, in the light-cone
formalism, $f_V$ is related to the leptonic decay width of the $J/\psi$ 
as $\Gamma(J/\psi \to e^+e^-)=(16\pi\alpha^2/27)|f_V|^2/M_{J/\psi}$. 
(See footnote 9 of BC.) Therefore, the value of $f_V$ is
independent of the shape of light-cone distribution and, hence, is
unaffected by the treatment of the high-momentum tail of the light-cone
distribution. The value of $f_P$ is fixed by making use of the fact that
$f_P$ is equal to $f_V$, up to corrections of relative order $v^2$, which
are neglected.

\subsection{Numerical results}
Now we compute numerical values of the light-cone cross section
(\ref{light-cone-x-section}), using the light-cone distribution 
amplitudes that we derived from the Cornell potential model. 

As we have already mentioned, we drop contributions that arise from the 
light-cone distribution amplitude $V_A$, as these are of higher order in $v$. 
That is, we set the term $f_5$ in Eq.~(\ref{fi}) to zero. For the model 
light-cone distribution amplitudes of BC, this has the effect of reducing the 
cross section from the value $\sigma(e^+e^-\to J/\psi+\eta_c)
\approx 33$~fb reported in BC to $\sigma(e^+e^-\to J/\psi+\eta_c)        
\approx 23.47$~fb. Thus, we do not expect the absence of this contribution 
to affect the order-of-magnitude discrepancy between the light-cone and 
NRQCD factorization results that we wish to address here.

The remaining light-cone distribution amplitudes are given in terms of the
light-cone distribution amplitude that is derived from the potential-model wave
function [Eq.~(\ref{lc-dists})]. Hence, we take 
\begin{subequations}
\label{VP}
\begin{eqnarray}
&&V_T(x)=V_L(x)=V_\perp(x)=\phi(x_1),
\\
&&P_P(y)=P_A(y)=\phi(y_1),
\end{eqnarray}
\end{subequations}
where $\phi(z)$ is defined in Eq.~(\ref{eq:phi-z-exact}), and we
use the equivalent forms in Eqs.~(\ref{eq:phib-z-exact}) and
(\ref{eq:phia-z-single}) for the numerical evaluation. Substituting
these light-cone distribution amplitudes into Eq.~(\ref{fi}), we obtain
the values for the cross section $\sigma(e^+e^-\to J/\psi+\eta_c)$ that
are shown in the fourth row of Table~\ref{table:sigma-phi1} and are
denoted by $\sigma_0$ .
 \begin{table}
 \caption{\label{table:sigma-phi1}
Values for the cross section $\sigma(e^+e^-\to J/\psi+\eta_c)$ computed
using the light-cone distribution amplitudes that are derived from a
potential-model wave function. As is explained in the text, $\sigma_0$
denotes the unsubtracted cross section, $\sigma$ denotes the cross
section in which contributions that correspond to order-$\alpha_s$
correction to the NRQCD short-distance coefficient have been removed
from the light-cone distribution amplitudes, and $\sigma_\delta$ denotes
the cross section in which the light-cone distribution amplitudes have been
replaced by $\delta$ functions. The notation ``BC'' indicates that the
values of the renormalization constants $Z_i$ or $\alpha_s$ are 
those that were used in BC. The value $\bar{M}_c=1.2$~GeV was used in BC; 
the values $\bar{M}_c=1.4$~GeV and $\alpha_s=0.21$ were used in
Ref.~\cite{Braaten:2002fi}. The notation ``$\overline{\rm BC}$'' indicates
that, in Eq.~(\ref{s}), we have inserted two factors of $Z_m^k$ for each
factor of ${\bar{M}}_Q^2$, as is described in the text below
Eq.~(\ref{s}).
}
 \begin{ruledtabular}
 \begin{tabular}{c|ccccc}
$Z$&                BC &$\overline{\rm BC}$ &1  &1    &1\\
$\alpha_s$&         BC &BC &BC &0.21 &0.21\\
$\bar{M}_c$~(GeV)&  1.2&1.2&1.2&1.2  &1.4\\
\hline
$\sigma_0$~(fb)&22.51&25.03&11.59& 7.58 & 4.95\\
$\sigma$~(fb)  &$\,\,\,$7.77&$\,\,\,$8.19&$\,\,\,$4.30& 3.29 & 2.48\\
$\sigma_\delta$~(fb)&\,\,\,$5.39$&$\,\,\,$5.59&$\,\,\,$3.04&2.48&1.95
 \end{tabular}
 \end{ruledtabular}
 \end{table}
If we correct the light-cone distribution amplitudes by subtracting the piece
$\Delta\phi$ [Eq.~(\ref{dphi-exact})] that corresponds to the 
order-$\alpha_s$ correction to the NRQCD short-distance coefficient, 
then we obtain the results that are shown in the fifth row of
Table~\ref{table:sigma-phi1} and are denoted by $\sigma$. Finally, if we
replace the light-cone distribution amplitudes $\phi(z)$ by $\delta(z-1/2)$, 
then we obtain the results that are shown in the sixth row
Table~\ref{table:sigma-phi1} and are denoted by $\sigma_{\delta}$ .

\subsection{Discussion}
\label{sec:discussion}

As can be seen from Table~\ref{table:sigma-phi1}, if we take the values
of $Z_i$, $\alpha_s$, and $\bar{M}_c$ that were used in BC, then 
the value of the unsubtracted cross section $\sigma$ that is obtained 
by using the potential-model light-cone distribution amplitude does 
not differ substantially from the value $23.47$~fb that is obtained 
by using the model light-cone distribution amplitudes of BC (with the
contribution of $V_A$ set to zero). This is not surprising, since, 
as can be seen from Fig.~\ref{fig:BC-Pot-dists}, the potential-model
light-cone distribution amplitude does not differ greatly in shape 
from the model light-cone distribution amplitudes of BC.

However, once the light-cone distribution amplitudes are corrected by 
subtracting the part $\Delta\phi$ that corresponds to the order-$\alpha_s$
correction to the NRQCD short-distance coefficient, then the cross
section is reduced significantly---by a factor of approximately three if
the values of $Z_i$, $\alpha_s$, and $\bar{M}_c$ are those that 
were used in BC.

A further large reduction in the cross section occurs if we set the
renormalization factors $Z_i$ equal to unity. These factors
represent an attempt to resum large logarithms of the heavy-quark 
virtuality in the production process divided by $\bar{M}_c^2$.
However, it is not clear to us that these factors, as employed in BC,
correctly resum such logarithms. The factors $Z_p$ and $Z_t$ are
inserted in an attempt to account for the evolution of the light-cone
distributions from the scale $\bar{M}_c^2$ to the scale of the
heavy-quark virtuality. However, these renormalization factors account
correctly only for the evolution of the first moments of those
distributions. The factor $Z_m$ evolves the heavy-quark mass in the
heavy-quark propagator. However, some of the factors of ${\bar{M}}_Q$ in
Eqs.~(\ref{d}) and (\ref{s}) should not be evolved because they do not
arise from the propagator mass, but, rather, from the approximation that
the external heavy-quark lines are taken to be on shell with mass
${\bar{M}}_Q$. Specifically, some of the factors of ${\bar{M}}_Q$ that
appear in the quantities $d(x,y)$, $s(x)$, and $s(y)$ [Eqs.~(\ref{d})
and (\ref{s})] arise in this way.

After we have set the renormalization factors $Z_i$ equal to unity,
the cross section is reduced to $4.3$~fb, which is
comparable to the values 
$3.78\pm 1.26~\hbox{fb}$ (Ref.~\cite{Braaten:2002fi}) and 
$5.5~\hbox{fb}$ (Ref.~\cite{Liu:2002wq})
that were obtained in the NRQCD factorization calculations in leading
order in $\alpha_s$. For purposes of comparison with the results of
Ref.~\cite{Braaten:2002fi}, we show in the last two columns of
Table~\ref{table:sigma-phi1} the effects of taking the values of
$\bar{M}_c$ and $\alpha_s$ that were used in that paper. It should be
noted in making this comparison that the NLO value of the wave function
at the origin that was used in Ref.~\cite{Braaten:2002fi} corresponds to
$f_V\approx 489$~MeV. Use of this value for $f_V$ (rather than
$400$~MeV) would enhance the cross section by about a factor $2.2$.

In BC, it is claimed that a large enhancement of the cross section
occurs in the light-cone calculation because it takes into account the
finite width of the quarkonium wave function. However, we see from
Table~\ref{table:sigma-phi1} that once the $\Delta\phi$ part of the
light-cone distribution amplitudes has been subtracted, the cross 
section does not differ greatly from that which is obtained by using 
$\delta$-function light-cone distribution amplitudes. We conclude that 
any large enhancement that arises from the finite width of the 
quarkonium wave function must be
coming from the region that corresponds to the order-$\alpha_s$
corrections to the NRQCD short-distance coefficient, {\it i.e.} the
region of large heavy-quark momentum ($\agt m_c$). Within our
potential-model calculation, it is essential to remove this region, as
it is not treated reliably under the nonrelativistic approximations that
go into that model. In comparing with NRQCD factorization calculations,
it is necessary to identify the contribution from this region, even 
if one makes use of light-cone distribution amplitudes with the 
correct high-momentum behavior, because otherwise its contributions 
would be double counted in the NRQCD factorization corrections of 
order $\alpha_s$ to the production cross section. We note 
that the order-$\alpha_s$ corrections to the production cross section 
are indeed large \cite{Zhang:2005ch}.

The absence of a large enhancement of $\sigma$ relative to
$\sigma_\delta$ is somewhat surprising, since it is known that, in the
NRQCD factorization approach, there are large corrections of
order $v^2$ (Ref.~\cite{Braaten:2002fi}) that account partially
for the effects of the finite width of the quarkonium wave function.
However, the order-$v^2$ corrections arise from several sources:
the dependence of the amplitude on the $Q\bar Q$ relative momentum and 
the difference between the charmonium mass and $2m_c$ in the phase
space and in the relativistic normalizations of the states. 
Of these corrections, only the first type is sensitive to the 
finite width of the quarkonium wave function. It is 
$3.87\langle v^2\rangle$ relative to the lowest-order cross
section \cite{Braaten:2002fi}. Here, $\langle v^2\rangle=\langle{\cal
P}_1\rangle/(m_c^2\langle{\cal O}_1\rangle )$, where the matrix 
elements are taken in the $J/\psi$ or $\eta_c$ state, and the NRQCD 
operators ${\cal O}_1$ and ${\cal P}_1$ are defined in 
Ref.~\cite{Bodwin:1994jh}.
(Their matrix elements are proportional to the square of the wave
function at the origin and the wave function at the origin times the
second derivative of the wave function at the origin, respectively.)
Taking $\langle{\cal P}_1\rangle/\langle{\cal O}_1\rangle=0.50\pm
0.09\pm 0.15$~GeV$^2$ (Ref.~\cite{BKL}) and $m_c=1.4$~GeV, we see that
the order-$v^2$ corrections that arise from the dependence of the
amplitude on the $Q\bar Q$ relative momentum are approximately
$99\%$ of the lowest-order cross section. The difference between
$\sigma$ and $\sigma_\delta$ is only approximately $29\%$. However,
it should be remembered that the light-cone formalism omits
contributions that arise from the transverse part of the $Q\bar Q$
relative momentum. (Note, though, that the transverse-momentum
contributions are suppressed by a factor $v^2M_{\rm meson}^2/E_{\rm
beam}^2$ relative to the longitudinal-momentum contributions.) We
also point out that, in dropping terms that are proportional to
the light-cone distribution amplitude $V_A$, we have omitted
corrections of order $v^2$ that, in the case of the model light-cone
distributions of BC, amount to about a $30\%$ correction relative to
the cross section computed without them. In order to be more
quantitative about the effects of these omissions, we have calculated the
order-$v^2$ correction that is associated with the finite widths 
of the light-cone distributions by expanding the light-cone cross section
around $z=1/2$. The result is a relative correction $2.1 \langle v^2
\rangle\approx 54\%$.
Presumably, the remaining small difference between this correction 
and $(\sigma-\sigma_\delta)/\sigma$ of about $54\%- 29\%= 25\%$ 
arises from corrections of still higher orders in $v$ and from the 
fact that we have computed $\delta\phi$ only at leading order in 
$\alpha_s$.

\section{Summary}
\label{sec:summary}

The discrepancy between theoretical predictions in the NRQCD
factorization approach and experimental measurements of the rate for the
process $e^+e^-\to J/\psi +\eta_c$ at the $B$ factories presents a
significant challenge to our understanding of the physics of quarkonia
and, perhaps, to our understanding of QCD. In Ref.~\cite{Bondar:2004sv}
(BC), it was suggested that the solution to this problem may lie in
using the light-cone formalism to take into account the effects of the
relative momenta of the heavy quarks and antiquarks in the 
quarkonia. The cross section calculated in BC is about a factor 
$6$--$10$ times larger than the cross sections sections calculated 
at leading order in the NRQCD factorization approach 
\cite{Braaten:2002fi,Liu:2002wq}. Both the NRQCD
factorization approach and the light-cone method are believed to
approximate QCD in the limit in which the hard-scattering momentum
transfer is much larger than the heavy-quark mass or $\Lambda_{\rm
QCD}$. Therefore, the apparent discrepancy between the two approaches
presents a further, purely theoretical, puzzle. In this paper, we
attempt to address this theoretical issue on several fronts.

First, we use a potential model to construct light-cone quarkonium
distribution amplitudes that are a good approximation to the true
light-cone distribution amplitudes of QCD. If one knows the static
heavy-quark potential sufficiently accurately, then the potential-model
allows one to determine the quarkonium wave function up to corrections
of relative order $v^2$, where $v$ is the heavy-quark (or antiquark)
velocity in the quarkonium rest frame. ($v^2\approx 0.3$ for
charmonium). We make use of the Cornell potential \cite{Eichten:1978tg},
which, for an appropriate choice of parameters, yields a good fit to the
lattice data for the heavy-quark potential---thus giving us confidence
in the connection of the potential model to QCD. We use the
Bethe-Salpeter equation to relate the potential-model wave functions to
the definitions of the light-cone distributions, which are written in
terms of the Bethe-Salpeter wave function. Then, we use on-shell
kinematics for the heavy quarks to relate momentum variables to
light-cone variables. The latter approximation is valid up to
corrections of relative order $v^2$. Finally, we integrate out the
relative $Q\bar Q$ transverse momentum in the Bethe-Salpeter wave
functions to form the light-cone distribution amplitudes. The light-cone
distribution amplitudes that we derive agree reasonably well in shape
with the model light-cone distribution amplitudes of BC. The calculated 
cross section changes very little if we replace the light-cone 
distribution amplitudes of BC with ours. Thus, we conclude that the
distribution amplitudes themselves are not the source of the theoretical
discrepancy.

A part of the light-cone result for the production cross section arises
from the region of large $Q\bar Q$ relative momentum ($\agt m_c$). This
region is not treated accurately in our inherently nonrelativistic
potential model. Furthermore, it consists of  contributions that are
part of the order-$\alpha_s$ correction to the hard-scattering
coefficient in the NRQCD factorization approach. Hence, it is necessary
to identify these contributions in order to avoid double-counting the
order-$\alpha_s$ correction in making comparisons between the light-cone
approach and the NRQCD factorization approach. After we subtract these
contributions from the light-cone calculation, the production cross
section is reduced by about a factor of three. 

The light-cone cross section in BC contains renormalization factors
$Z_i$ for the heavy-quark mass and the pseudo-scalar and tensor
point-like vertices. If we set them to unity, a further reduction in the
cross section of about a factor of two occurs. These renormalization
factors have no counterpart in a conventional NRQCD factorization
calculation. They represent an attempt to resum large logarithms of
the heavy-quark virtuality in the production process divided by the
square of the heavy-quark mass. However, as we have explained in
Sec.~\ref{sec:discussion}, we are not convinced that these
renormalization factors, as employed in BC, resum such logarithms
correctly. It is important to understand whether a correct resummation
yields a significant enhancement to the cross section and, if so, to
incorporate it into NRQCD factorization calculations, as well as into
light-cone calculations.

Once the factors $Z_i$ have been set equal to unity, the
subtracted light-cone cross section is comparable in size with the
NRQCD factorization cross sections, and it differs by only about 
$29\%$ from the light-cone cross section that is obtained by 
replacing the light-cone distribution amplitudes with zero-width distribution
amplitudes ($\delta$~functions).\footnote{Even in the case of
$\delta$-function light-cone distribution amplitudes, there are small
differences between the light-cone and NRQCD factorization cross
sections. Some of these differences arise from different choices for the
quarkonium wave functions at the origin, $\alpha_s$, and the charm-quark
mass and from the inclusion of quantum-electrodynamic corrections
in Ref.~\cite{Braaten:2002fi}. Other differences, presumably, are due to
the fact that the light-cone and NRQCD factorization cross sections
become equal only in the limit of infinite hard-scattering momentum
transfer.}

We conclude that there is no conflict between the light-cone and NRQCD
factorization results. Further, it seems that enhancements to the cross
section that arise from making use of a light-cone wave function of
finite width are rather modest and are not sufficient to remove the
discrepancy between theory and experiment. That discrepancy remains an
important puzzle. At this point, the most promising avenue for its
resolution seems to be the inclusion of corrections of higher order in
$\alpha_s$ (Ref.\cite{Zhang:2005ch}) and $v$.

\begin{acknowledgments}
We wish to thank Eric Braaten for useful discussions.
J.L.\ thanks the high energy theory group at Argonne National Laboratory
for its hospitality.
The research of G.T.B.\ in the High Energy Physics Division at
Argonne National Laboratory is supported by the U.~S.~Department of
Energy, Division of High Energy Physics, under Contract W-31-109-ENG-38.
The work of J.L.\ is supported by the Korea Research Foundation
under grant KRF-2004-015-C00092.
D.K.'s work is supported in part by the Seoul Science Fellowship 
and by the Korea Research Foundation under grant KRF-2006-612-C00003.
\end{acknowledgments}

\appendix
\section{Light-cone distribution amplitudes for the pure Coulomb potential and 
asymptotic behavior at $\bm{z=0,1}$}
\label{App:Coulomb}%

In this Appendix, we compute analytically the light-cone distribution 
amplitudes
$\phi$, $\phi_a$, and $\phi_b$ for the special case of a pure Coulomb
potential. We also determine their asymptotic behaviors at $z=0$ (or
$z=1$). These analytic results are useful in checking the accuracy of
the numerical-integration methods that we use in the more general case
of the Cornell potential.

We drop the linear potential in the radial equation (\ref{u-eq}).
In this pure Coulomb case, as is well known, Eq.~(\ref{u-eq}) can be
solved analytically. For the ground state, the wave function $u(\rho)$
and eigenvalue $\zeta$ are
\begin{eqnarray}
u(\rho)&=&\sqrt{\frac{\lambda^3}{2}}\rho e^{-\frac{\lambda}{2}\rho},
\\
\zeta_{10}&=&-\frac{\lambda^2}{4}.
\end{eqnarray}
Applying the relations in Eqs.~(\ref{kappa}), (\ref{kappa-lambda}), 
(\ref{r-rho}), and (\ref{eq:psir}),
we obtain 
\begin{subequations}
\begin{eqnarray}
R(r)&=&2 \gamma_{\textrm{C}}^{3/2} e^{-\gamma_{\textrm{C}} r},
\label{R-Coulomb}\\
\epsilon_{\textrm{B}}&=&-\frac{\gamma^2_{\textrm{C}}}{m}.
\end{eqnarray}
\end{subequations}
From Eq.~(\ref{R-Coulomb}), it follows that
\begin{equation}
\left[\psi(r)\right]_{\textrm{Coulomb}}
=\psi(0)e^{-\gamma_{\textrm{C}} r},
\label{psi-coulomb}
\end{equation}
where the wave function at the origin is given by
\begin{equation}
|\psi(0)|=\sqrt{\frac{ \gamma^3_{\textrm{C}} }
                   {\pi}}.
\end{equation}

We next use Eq.~(\ref{psi-coulomb}) to calculate the light-cone
distribution amplitude in the pure Coulomb case. Substituting 
Eq.~(\ref{psi-coulomb}) into Eq.~(\ref{eq:phiab-z-exact})
and integrating over $r$, we obtain
\begin{subequations}
\label{phi-coulomb-kint}
\begin{eqnarray}
\label{phi-coulomb-kint-full}
\left[\phi(z)\right]_{\textrm{Coulomb}}&=&
\frac{4\gamma_{\textrm{C}}}{\pi}
\int_{\sqrt{d(z)}}^\infty
\frac{|\bm{p}|\sqrt{|\bm{p}|^2+m_c^2}}{(|\bm{p}|^2
+\gamma_{\textrm{C}}^2)^2}\,d|\bm{p}|,
\\
\left[\phi_a(z)\right]_{\textrm{Coulomb}}&=&
\frac{4\gamma_{\textrm{C}}}{\pi}
\int_{\sqrt{d(z)}}^\infty
\frac{|\bm{p}|^2}{(|\bm{p}|^2+\gamma_{\textrm{C}}^2)^2}\,d|\bm{p}|,
\end{eqnarray}
\end{subequations}
where we have used $\phi=\phi_a+\phi_b$. Carrying out the integration
over $|\bm{p}|$ in Eqs.~(\ref{phi-coulomb-kint}), we have
\begin{subequations}
\begin{eqnarray}
\left[\phi(z)\right]_{\textrm{Coulomb}}
&=&
\frac{2\gamma_{\textrm{C}}}{\pi}
\left[
\frac{\sqrt{d(z)+m_c^2}
     }{d(z)+\gamma_{\textrm{C}}^2}
+
\frac{\sinh^{-1}
 \sqrt{\frac{m_c^2-\gamma_{\textrm{C}}^2
            }{d(z)+\gamma_{\textrm{C}}^2}
      }
     }
 {\sqrt{m_c^2-\gamma_{\textrm{C}}^2}}
\right],
\label{phi-z-coulomb}
\\
\left[\phi_a(z)\right]_{\textrm{Coulomb}}
&=&
\frac{2}{\pi}
\left[
\frac{\gamma_{\textrm{C}} \sqrt{d(z)}}{d(z)+\gamma_{\textrm{C}}^2}
+\arctan\left(\frac{\gamma_{\textrm{C}}}{\sqrt{d(z)}}\right)
\right].
\label{phia-z-coulomb}
\end{eqnarray}
\end{subequations}
Note that Eq.~(\ref{phia-z-coulomb}) can also be obtained by substituting
the wave function in Eq.~(\ref{psi-coulomb}) into 
Eq.~(\ref{eq:phia-z-single}) and integrating over $r$.

We can use our results in the pure Coulomb case to compute the
asymptotic behavior of the light-cone distribution amplitudes at $z=0$
(or $z=1$). At $z=0$, the corresponding longitudinal momentum is infinite
[Eq.~(\ref{eq:k3})], and only the short-distance behavior of $\psi(r)$
is important. That short-distance behavior is governed by the Coulomb
potential.\footnote{By differentiating Eqs.~(\ref{eq:phia-z-exact}) and
(\ref{eq:phib-z-exact}) with respect to $z$ and setting $z=0$, one can
verify that, at $z=0$, only the large $p$ behaviors of the integrands
for $\phi_a$ and $\phi_b$ are important and, hence, that only the
short-distance behavior of $\psi(r)$ is important. This approach yields
a simple expression for the asymptotic behaviors of the derivatives of
the light-cone distribution amplitudes with respect to $z$, which can
then be translated into the asymptotic behaviors of the distribution
amplitude.} Taking the limit $z\to 0$ in Eqs.~(\ref{phi-z-coulomb}) and
(\ref{phia-z-coulomb}) and using $\phi_b=\phi-\phi_a$, we find the
asymptotic behaviors
\begin{subequations}
\label{asymptotics}
\begin{eqnarray}
\phi(z)&\sim&\phi_a(z)\sim \frac{8\gamma_C}{\pi m_c}z^{1/2},
\label{asympt-phi}\\
\phi_b(z)&\sim&\frac{16\gamma_C}{3\pi m_c}z^{3/2}.
\end{eqnarray}
\end{subequations}

The asymptotic forms in Eqs.~(\ref{asymptotics}) provide good tests of
the convergence in the high-momentum region of the numerical 
integrations that we use to compute $\phi(z)$ and $\phi_a(z)$.



\begin{thebibliography}{}

\bibitem{Pakhlov:2004au}                                                
  P.~Pakhlov  [Belle Collaboration],                                    
  arXiv:hep-ex/0412041.              

\bibitem{Abe:2004ww}
  K.~Abe {\it et al.}  [Belle Collaboration],
  Phys.\ Rev.\ D {\bf 70}, 071102 (2004)
  [arXiv:hep-ex/0407009].

\bibitem{Aubert:2005tj}
  B.~Aubert {\it et al.}  [BABAR Collaboration],
  Phys.\ Rev.\ D {\bf 72}, 031101 (2005)
  [arXiv:hep-ex/0506062].

\bibitem{Braaten:2002fi}
E.~Braaten and J.~Lee,
Phys.\ Rev.\ D {\bf 67}, 054007 (2003)
  [Erratum-ibid.\ D {\bf 72}, 099901 (2005)]
  [arXiv:hep-ph/0211085].

\bibitem{Liu:2002wq}
  K.~Y.~Liu, Z.~G.~He, and K.~T.~Chao,
  Phys.\ Lett.\ B {\bf 557}, 45 (2003)
  [arXiv:hep-ph/0211181].

\bibitem{Bodwin:1994jh}
  G.~T.~Bodwin, E.~Braaten, and G.~P.~Lepage,
  Phys.\ Rev.\ D {\bf 51}, 1125 (1995)
  [Erratum-ibid.\ D {\bf 55}, 5853 (1997)]
  [arXiv:hep-ph/9407339].

\bibitem{Zhang:2005ch}
  Y.~J.~Zhang, Y.~j.~Gao, and K.~T.~Chao,
  Phys.\ Rev.\ Lett.\  {\bf 96}, 092001 (2006)
  [arXiv:hep-ph/0506076].
\bibitem{Bondar:2004sv}

  A.~E.~Bondar and V.~L.~Chernyak,
  Phys.\ Lett.\ B {\bf 612}, 215 (2005)
  [arXiv:hep-ph/0412335].

\bibitem{Ma:2004qf}
  J.~P.~Ma and Z.~G.~Si,
  Phys.\ Rev.\ D {\bf 70}, 074007 (2004)
  [arXiv:hep-ph/0405111].

\bibitem{Braguta:2005kr}
  V.~V.~Braguta, A.~K.~Likhoded, and A.~V.~Luchinsky,
  Phys.\ Rev.\ D {\bf 72}, 074019 (2005)
  [arXiv:hep-ph/0507275].

\bibitem{Braguta:2006nf}
  V.~V.~Braguta, A.~K.~Likhoded, and A.~V.~Luchinsky,
  Phys.\ Lett.\ B {\bf 635}, 299 (2006)
  [arXiv:hep-ph/0602047].

\bibitem{Nayak:2005rw}
  G.~C.~Nayak, J.~W.~Qiu, and G.~Sterman,
  Phys.\ Lett.\ B {\bf 613}, 45 (2005)
  [arXiv:hep-ph/0501235].

\bibitem{Nayak:2005rt}
  G.~C.~Nayak, J.~W.~Qiu, and G.~Sterman,
  Phys.\ Rev.\ D {\bf 72}, 114012 (2005)
  [arXiv:hep-ph/0509021].

\bibitem{Bodwin:2005ec}
  G.~T.~Bodwin,
  Int.\ J.\ Mod.\ Phys.\ A {\bf 21}, 785 (2006)
  [arXiv:hep-ph/0509203].

\bibitem{Nayak:2006fm}
  G.~C.~Nayak, J.~W.~Qiu and G.~Sterman,
  Phys.\ Rev.\ D {\bf 74}, 074007 (2006)
  [arXiv:hep-ph/0608066].

\bibitem{Brambilla:1999xf}
  N.~Brambilla, A.~Pineda, J.~Soto, and A.~Vairo,
  Nucl.\ Phys.\ B {\bf 566}, 275 (2000)
  [arXiv:hep-ph/9907240].

\bibitem{Eichten:1978tg}
  E.~Eichten, K.~Gottfried, T.~Kinoshita, K.~D.~Lane, and T.~M.~Yan,
  Phys.\ Rev.\ D {\bf 17}, 3090 (1978)
  [Erratum-ibid.\ D {\bf 21}, 313 (1980)].

\bibitem{Bali:2000gf}
  G.~S.~Bali,
  Phys.\ Rept.\  {\bf 343}, 1 (2001)
  [arXiv:hep-ph/0001312].

 \bibitem{BKL}
  G.~T.~Bodwin, D.~Kang, and J.~Lee,
  Phys.\ Rev.\ D {\bf 74}, 014014 (2006)
  [arXiv:hep-ph/0603186].

\bibitem{Booth:1992bm}
  S.~P.~Booth, D.~S.~Henty, A.~Hulsebos, A.~C.~Irving, C.~Michael, and 
P.~W.~Stephenson
                  [UKQCD Collaboration],
  Phys.\ Lett.\ B {\bf 294}, 385 (1992)
  [arXiv:hep-lat/9209008].

\bibitem{Gupta:1996sa}
  R.~Gupta and T.~Bhattacharya,
   Phys.\ Rev.\ D {\bf 55}, 7203 (1997)
     [arXiv:hep-lat/9605039].

\bibitem{Kim:1993gc}
  S.~Kim and D.~K.~Sinclair,
  Phys.\ Rev.\ D {\bf 48}, 4408 (1993).

\bibitem{Kim:1996cz}                                          
  S.~Kim and S.~Ohta,                                                    
  Nucl.\ Phys.\ Proc.\ Suppl.\  {\bf 53}, 199 (1997)                     
  [arXiv:hep-lat/9609023].                            

\bibitem{Chernyak:1983ej}
  V.~L.~Chernyak and A.~R.~Zhitnitsky,
  Phys.\ Rept.\  {\bf 112}, 173 (1984).

\bibitem{Kuhn:1979bb}                                                  
J.~H.~K\"uhn, J.~Kaplan, and E.~G.~Safiani,
Nucl.\ Phys.\ {\bf B157}, 125 (1979).
 
\bibitem{Guberina:1980dc}
B.~Guberina, J.~H.~K\"uhn, R.~D.~Peccei, and R.~R\"uckl,
Nucl.\ Phys.\ {\bf B174}, 317 (1980).

\bibitem{Bodwin:2002hg}
  G.~T.~Bodwin and A.~Petrelli,
  Phys.\ Rev.\ D {\bf 66}, 094011 (2002)
  [arXiv:hep-ph/0205210].

\bibitem{Salpeter:1951sz}                                     
  E.~E.~Salpeter and H.~A.~Bethe,                                      
  Phys.\ Rev.\  {\bf 84}, 1232 (1951).                                        

\bibitem{Salpeter:1952ib}
  E.~E.~Salpeter,
  Phys.\ Rev.\  {\bf 87}, 328 (1952).

\bibitem{Beneke:1997zp}
  M.~Beneke and V.~A.~Smirnov,
  Nucl.\ Phys.\ B {\bf 522}, 321 (1998)
  [arXiv:hep-ph/9711391].

\bibitem{Brodsky:1981kj}
S.~J.~Brodsky and G.~P.~Lepage,
Phys.\ Rev.\ D {\bf 24}, 2848 (1981).


\end{thebibliography}
\end{document}